\documentclass[twocolumn,showpacs,preprintnumbers,nofootinbib,amsmath,amssymb]{revtex4}
\font\mybb=msbm10 at 10pt
\def\bb#1{\hbox{\mybb#1}}
\begin{document}

\title{Covariant action and equations of motion for the eleven dimensional multiple
M$0$-brane system}

\author{Igor A. Bandos $^{\dagger\ddagger}$ and Carlos Meliveo $^{\dagger}$
}
\address{$^{\dagger}$Department of
Theoretical Physics, University of the Basque Country UPV/EHU,
P.O. Box 644, 48080 Bilbao, Spain
 \\ $^{\ddagger}$
IKERBASQUE, Basque Foundation for Science, 48011, Bilbao, Spain}

\date{April 1, 2013. V2: June 20, 2013. Printed \today}

\def\theequation{\arabic{section}.\arabic{equation}}

\begin{abstract}

We study the properties of the covariant supersymmetric and $\kappa$--symmetric action
for a system of N nearly coincident  M$0$-branes (mM0 system) in flat eleven
dimensional (11D) superspace and obtain supersymmetric equations for this dynamical
system. Although a single M$0$ brane is the  massless 11D superparticle, center of
energy motion of the mM0 system is characterized by a nonnegative constant mass $M$
constructed from the matrix fields describing the relative motion of mM$0$
constituents.  We show that a bosonic solution of the mM0 equations can be
supersymmetric iff this effective mass vanishes, $M^2=0$, and that all the
supersymmetric bosonic solutions preserve just one half of the 11D supersymmetry.

\end{abstract}

\pacs{
11.25.-w, 11.25.Yb, 04.65.+e, 11.10.Kk, 11.30.Pb}

\maketitle

\section{Introduction}

In \cite{Witten:1995im} it was motivated that an approximate description of the system
of nearly coincident Dirichlet $p$--branes (D$p$-branes) is provided by maximal $d=p+1$
dimensional supersymmetric Yang--Mills  (SYM) theory with the gauge group $U(N)$, which
can be obtained by the dimensional reduction of D=10 U(N) SYM theory. This includes
$D-p-1$ Hermitian matrices of scalar fields the diagonal elements of which describe the
positions of different D$p$-branes while the off--diagonal elements account for the
strings stretched between different D$p$-branes.

As a single D$p$-brane was known to be described by a sum of supersymmetric
Dirac--Born--Infeld action, providing a nonlinear generalization of the U(1)
Yang--Mills action, and a Wess--Zumino term (see \cite{B+T=Dpac} and refs therein), it
was natural to search for a nonlinear generalization of the non--Abelian SYM action
providing a more complete nonlinear description of the system of nearly coincident
D$p$-branes. For the bosonic limit of multiple nearly coincident D$p$-branes (mD$p$
system) the most popular description is provided by Myers 'dielectric brane' action
\cite{Myers:1999ps}. This was obtained by a chain of T-duality transformations from the
10D symmetric trace non-Abelian Born--Infeld action, proposed by Tseytlin
\cite{Tseytlin:DBInA} for purely bosonic limit of the system of multiple  spacetime
filling D$9$--branes (mD$9$ system). Both the  actions of \cite{Tseytlin:DBInA} and
\cite{Myers:1999ps} resisted the attempts to construct their supersymmetric
generalizations for many years; in addition the Myers action does not possess the
Lorentz symmetry.

The supersymmetric and Lorentz covariant description of the mD$p$ system was reached in
\cite{Howe+Linstrom+Linus=2005,Howe+Linstrom+Linus=2007,HLW2} in the frame of the so--called 'boundary fermion approach'.
However, this description  is provided at the 'minus one quantization level', which
means that, to reach the description of mD$p$ system similar to the description of
D$p$--branes in {\it e.g.} \cite{B+T=Dpac}, one has to perform quantization of the
dynamical system. This task is nontrivial and has not been solved in a complete form
\footnote{For the bosonic limit, {\it i.e.} when the boundary fermions are the only non-vanishing fermions, it was shown in  \cite{Howe+Linstrom+Linus=2007}  that replacing these by suitable Dirac matrices, replacing the Poisson brackets by (anti)commutators and replacing the integration over the boundary fermions by symmetric trace of the product of matrices one obtains  a result which agrees with that of \cite{Myers:1999ps}. The supersymmetrization of the action of \cite{Howe+Linstrom+Linus=2007} was constructed in  \cite{HLW2} with the use of Bernstein-Leites integration over the boundary fermion coordinates. It is invariant under $\kappa$--symmetry with parameters dependent on  the boundary fermions, which suggests it should be a matrix after quantization of the boundary fermion sector. Such a quantization of supersymmetric action of \cite{HLW2} has not been developed yet. },
which motivated further attempts to obtain a possibly  approximate but Lorentz
covariant and supersymmetric description of the mD$p$ system going beyond the SYM
approximation (see e.g. \cite{IB09:D0}). Only for the case of mD$0$--system a
nonlinear, supersymmetric and Lorentz invariant candidate for the wanted mD$0$ action
does exist \cite{Dima+=JHEP03}.

As D$p$--branes with $p=0,2,4$ can be obtained by a  dimensional reduction of the
11 dimensional  M$0$, M$2$ and M$5$ branes, it is natural to expect that the
corresponding mD$p$ system can be obtained from the respective mM$p$ system. However,
for the case of mM$5$ even the question on what should be a counterpart of the very low energy approximate SYM description is still obscure (see {\it e.g.} \cite{SSWW} for
related study and references). For the case of  mM$2$ brane such a problem was unsolved many years, but relatively recently the $d=3$ ${\cal N}=8$ supersymmetric BLG model
\cite{BLG} based on a 3-algebra (see \cite{deAzcarraga:2010mr} and refs. therein)
instead of Lie algebra, and then a more conventional  ABJM model \cite{ABJM}
(with $SU(N)\times SU(N)$ gauge symmetry and  only ${\cal N}=6$ manifest
supersymmetries) were found and accepted for this role.

As far as multiple M$0$ brane (mM$0$) system is considered, a purely bosonic candidate
was constructed in \cite{YLozano+=0207} as  the 11D generalization of the Myers
'dielectric D$0$-brane' action. On the other hand, an approximate but supersymmetric
and Lorentz covariant equations of motion for mM$0$--system in flat target 11D superspace were obtained  in
\cite{mM0=PLB} in the frame of superembedding approach (see \cite{bpstv,hs2} as well as \cite{Dima99,IB09:M-D} and refs. therein). The generalization of these equations for
the case of mM$0$--system in curved 11D supergravity superspace, which describes the
generalization of the M(atrix) theory  \cite{Banks:1996vh}(see also earlier \cite{de Wit:1988ig})
for the case of its interaction with arbitrary supergravity background, were presented and studied in
\cite{mM0=PRL}. In \cite{mM0-pp=PRD} it was shown that in the case of 11D pp-wave
background these equations reproduce (in an approximation) the so--called BMN  matrix
model proposed for this background by Berenstein, Maldacena and Nastase in \cite{BMN}.
This result, confirming that the equations of \cite{mM0=PRL} describe the Matrix theory interacting with supergravity background, also have shown that, due to the superspace
origin of these equations, their applications for a certain, even purely bosonic
supergravity background are sufficiently complicated: it requires the lifting of the
bosonic supersymmetric solution of the supergravity equations till the complete
superfield solution of the 11D superspace supergravity constraints \cite{BrinkHowe80}.
This made desirable to find an action which reproduces the Matrix model equations of
\cite{mM0=PRL} or their generalizations.

For the case of mM0 system in flat target superspace such an action was proposed in
\cite{mM0=action}, where it was shown that it possesses local ${\cal N}=16$ 1d
supersymmetry. The main aim of this paper is to derive and to study equations of motion of the mM0 system described by that action. We will study the properties of the supersymmetric solutions of these equations, show that their center of energy sector is similar to the solution of the equation for a single M$0$-brane, and also present two examples of non-supersymmetric solutions with different properties of the center of energy motion.
It was noticed  in \cite{mM0=action} that all the supersymmetric solutions of the mM0 equations are characterized by vanishing of the effective mass of the center of energy motion. Here we not only reproduce this result proving that $M^2=0$ appears as the BPS equation, but also show that all the supersymmetric solutions of mM0 equations preserve just 1/2 of the 11D supersymmetry, so that all the supersymmetric mM0 BPS states are 1/2 BPS.

The paper is organized as follows. In Secs. \ref{M0sec}, \ref{M0eqs} we review the spinor moving
frame formulation of single M$0$ brane, this is to say of 11D massless superparticle.
We describe there the moving frame and 11D spinor moving frame variables (sec.
\ref{M0sec}C,D  and also \ref{M0sec}E,F), discuss the M$0$ brane equations of motion (Sec. \ref{M0eqs})
obtained from the spinor moving frame action (of sec. \ref{IIA}) and show (in sec.
\ref{M0eqs}A) that supersymmetric solution of these equations preserve just 1/2 (16 of 32)  of the 11D supersymmetries  {\it i.e.} describe ${1\over 2}$ BPS states. We also discuss there the irreducible $\kappa$--symmetry of the spinor
moving frame formulation of superparticle (sec. \ref{sec=irrKap}),  stress its
identification with the local worldline supersymmetry\footnote{This was for the first
time found in \cite{stv} in simpler, D=3,4 superparticle models. } and describe (in Sec. \ref{M0eqs}) the composite 1d ${\cal N}=16$
supergravity multiplet corresponding to it. This supergravity induced by embedding of
the M$0$ worldline into the target 11D superspace allows to make local the originally global supersymmetry of, say, 1d ${\cal N}=16$ supersymmetric Yang--Mills (SYM) theory living on the worldline and plays an important role in the mM$0$
action of \cite{mM0=action}, which is the subject of our investigation here.  This action (described
in Sec. \ref{mM0secAC}) is given by the sum of the 1d ${\cal N}=16$ SYM action coupled to the induced 1d ${\cal N}=16$ supergravity and of the same  spinor moving
frame action functional as  we have used to describe the single M$0$, which now
describes the center of energy motion of the mM$0$ system. In Sec. V we vary this mM0 action and obtain the set of covariant and supersymmetric equations of motion for mM0 system. The  properties of the solutions of these equations are studied in Secs. VI, VII and VIII. Particularly, Sec. VII is devoted to supersymmetric solutions of mM$0$ equations. We show there that all these are characterized by vanishing center of energy effective mass, $M^2=0$. In Sec. VIII we  present two examples of non--supersymmetric solutions with $M^2\not= 0$. We conclude and discuss our results in Sec. IX. Appendices A and C contain the complete lists of the equations of motion for  single M0--brane and for the mM0 systems respectively. Appendix B collects the properties of the moving frame and spinor moving frame variables.

\section{Single M0--brane in spinor moving frame formulation}
\label{M0sec}
\setcounter{equation}{0}

In this section we review the spinor moving frame formulation of single M$0$-brane, this is to say 11D massless superparticle, developed in
\cite{IB07:M0}, and summarize the properties of spinor moving frame variables used in this formulation as well as in the description of multiple M$0$-brane (mM$0$) system.

\subsection{Twistor--like spinor moving frame action and its irreducible
$\kappa$--symmetry.}
\label{IIA}

The spinor moving frame action of M$0$--brane reads (see \cite{IB07:M0} and also
\cite{B90,IB+AN=95,BL98',IB+JdA+DS=2006} and \cite{BZ-str,bsv})
\begin{eqnarray}
\label{SM0=} S_{M0} &=& \int_{W^1} \rho^{\#}\, \hat{E}^{=} = \int_{W^1} \rho^{\#}\,
u_a^{=}  \, {E}^a(\hat{Z}) \qquad
\\ \label{SM0==}
& =& {1/16}\int_{W^1}\rho^{\#}\, (v_{q}^{\; -}{\Gamma}_a v_{q}^{\; -}) \, \hat{E}^{a}
\; .
  \end{eqnarray}
In the first line of this equation, (\ref{SM0=}), $\rho^{\#}(\tau)$  is a Lagrange
multiplier,
\begin{eqnarray}
\label{hEa=} \hat{E}^{a}:= {E}^a(\hat{Z})=d\hat{Z}{}^M(\tau)  {E}_M^{a}(\hat{Z})=:
d\tau \hat{E}_\tau^a(Z)\;
  \end{eqnarray}
is the pull--back of the bosonic supervielbein of the 11D target superspace
($a=0,1,...,10$), $E^a= E^a(Z)= dZ^ME_M^a(Z)$,  to the worldline $W^1$ parametrized by
proper time $\tau$. In the case of flat target superspace the supervielbein
can be chosen in the form \footnote{The action (\ref{SM0=}), (\ref{SM0==}) makes sense
when  supervielbein  $E^a= dZ^ME_M^a(Z)$ obeys the 11D superspace supergravity
constraints \cite{BrinkHowe80}. In this paper we will restrict ourselves by the
case of flat target superspace, described by Eqs. (\ref{Ea=Pi}).}
\begin{eqnarray}
\label{Ea=Pi} E^a = \Pi^a = dx^a - i d\theta \Gamma^a\theta\; , \quad
E^\alpha=d\theta^\alpha
 \;   \quad
  \end{eqnarray}
\footnote{We use the (real)  matrices   $\Gamma^a_{\alpha\beta}=\Gamma^a_{\beta\alpha}=
\Gamma^a_{\alpha}{}^\gamma C_{\gamma\beta}$ and
$\tilde{\Gamma}_a^{\alpha\beta}=\tilde{\Gamma}_a^{\beta\alpha}=
C^{\alpha\gamma}\Gamma^a_{\gamma}{}^{\beta}$  constructed as a product of  11D Dirac
matrices $\Gamma^a_{\beta}{}^\gamma$ (obeying $\Gamma^a\Gamma^b +
\Gamma^b\Gamma^a=2\eta^{ab} I_{32\times 32}$) with, respectively, the 11D charge
conjugation matrix  $C_{\gamma\beta}=- C_{\beta\gamma}$ and its inverse
$C^{\alpha\beta}=- C^{\beta\alpha}$. Both $\Gamma^a_{\beta}{}^\gamma$  and
$C_{\beta\gamma}$ are pure imaginary in our mostly minus notation $\eta^{ab}=diag
(1,-1, ..., -1)$.}. Finally, $\hat{E}^{=}= \hat{E}^{a}u_a^{=}$ and $u_a^{=}=u_a^{=}(\tau)$ is a
light--like 10D vector,  $u^{=a}u_a^{=}=0$.

One can write the action (\ref{SM0=}) in a probably more conventional from, extracting
$d\tau$ measure from the pull--back of the supervielbein 1--form (see (\ref{hEa=}))
\begin{eqnarray}
\label{SM0=dt} S_{M0} &=& \int_{W^1} d\tau \rho^{\#}\, \hat{E}_\tau^{=}= \qquad
\nonumber \\ &=&\int_{W^1} d\tau \rho^{\#}\, \partial_\tau \hat{Z}{}^M(\tau)
{E}_M^{a}(\hat{Z}(\tau)) u_a^=(\tau) \; .
  \end{eqnarray}
We however, prefer to hide $d\tau$ inside of differential form, define the
Lagrangian 1-form by ${\cal L}_1=d\tau {\cal L}_\tau$,  and write our actions as
integral of this 1--form over the worldline, $\int_{W^1} {\cal L}_1$, rather than as an integral
over $d\tau$ of a density,  $\int d\tau {\cal L}_\tau$.

If we were stoping at this stage, one can easily observe that the action (\ref{SM0=})
can be obtained from the first order form of 11D version of the Brink--Schwarz action,
\begin{eqnarray}
\label{S'M0=} S_{BS} &=& \int_{W^1} \left( p_a \hat{E}^{a} -{e\over 2} p_ap^a d\tau
\right) \; ,
  \end{eqnarray}
by solving the constraints  $p_ap^a=0$ (equations of motion for Lagrange multiplier
$e(\tau)$) and substituting them back to the action.  Furthermore, one might wonder why
the solution $p_a=\rho^{\#}u_a^=$ is written with a multiplier $\rho^{\#}(\tau)$ instead
of just stating that it has the form of $S= \int_{W^1}  p_a \hat{E}^{a}$ with $p_a$ constrained by
$p_ap^a=0$. We will answer that question a bit later, just announcing now that
$\rho^{\#}$ is a kind of St\"{u}ckelberg variable allowing to introduce an $SO(1,1)$
gauge symmetry; although looking artificial at this stage, this symmetry allows to
clarify the group theoretical meaning of $u_a^=$ and also of the set of $16$
constrained spinors appearing in the second representation of $S_{M0}$, Eq.
(\ref{SM0==}).

The light--like vector $u_a^=$ can be considered as a composite of (any of) the $16$
spinors $v^{-\alpha}_q$  provided these are constrained by
\begin{eqnarray} \label{Iu--=vGv}
 v_q^{-\alpha} (\Gamma^a)_{\alpha\beta} v_p^{-\beta}= \delta_{qp}
u^{=}_{ a} \;  \qquad && (\ref{Iu--=vGv}a)  \nonumber \\  2v_q^{-\alpha} v_q^{-\beta}= u^{=}_{ a}
\tilde{\Gamma}^{a\alpha\beta} \; . \qquad && (\ref{Iu--=vGv}b)
\end{eqnarray}
Notice that the traces of both equations give $16 u^{=}_{ a} = v_q^{-\alpha} (\Gamma^a)_{\alpha\beta} v_q^{-\beta}$ which can be read off (\ref{SM0==}) and (\ref{SM0=}). The set of spinors $v_q^{-\alpha}$ constrained by (\ref{Iu--=vGv}) are called {\it spinor moving frame variables} (hence the name `spinor moving frame' for the formulation of superparticle mechanics based on the action (\ref{SM0=}), (\ref{SM0==})).
Before discussing their  origin and nature (in sec. \ref{smovfr}) \cite{B90,Ghsds,BZ-str,GHT93,IB+AN=95}, in the next sec. \ref{sec=irrKap}   we would like to try to convince the reader in the usefulness of these 'square roots' of the light--like vector $u^{=}_{ a}$.

\subsection{Irreducible $\kappa$--symmetry of the spinor moving frame action}
\label{sec=irrKap}
The action (\ref{SM0=}), (\ref{SM0==}) is invariant under the following local fermionic $\kappa$--symmetry transformations
\begin{eqnarray}
\label{kap=irr} && \delta_\kappa \hat{x}^a =  - i  \hat{\theta}
\Gamma^a\delta_\kappa\hat{\theta}\; , \qquad \delta_\kappa \hat{\theta}^\alpha =
\epsilon^{+q} (\tau)  v_q^{-\alpha} \; , \quad \nonumber \\ && \delta_\kappa
\rho^{\#}=0 \; , \quad \nonumber \\   && \delta_\kappa u_a^{=}=0 \quad \Leftarrow \quad
\delta_\kappa v_q^{-\alpha}=0
 \; .  \quad
\end{eqnarray}
These symmetry is {\it irreducible} in the sense of that each of 16 fermionic parameters\footnote{The
$\kappa$--symmetry was discovered in \cite{kappaAL,kappaS}  and was shown to coincide with the local worldline supersymmetry in \cite{stv}. Our notation
$\epsilon^{+q} (\tau) $ for the (irreducible)  $\kappa$--symmetry parameter is an implicit reference on this later result which will be useful in the discussion below.}
 $\;\epsilon^{+q} (\tau) $ acts efficiently on the variables of the theory and can be used to remove some components of fermionic field
$\hat{\theta}^\alpha(\tau)$ thus reducing the number of the degrees of freedom in it to $16$ (while $\alpha=1,...,32$).

In contrast, the $\kappa$--symmetry of the original Brink--Schwarz superparticle action
(\ref{S'M0=})  \cite{kappaS}
\begin{eqnarray}
\label{kappa} \delta_\kappa \hat{x}^a =  - i  \hat{\theta}
\Gamma^a\delta_\kappa\hat{\theta}\; , \qquad \delta_\kappa \hat{\theta}^\alpha=p_a
\tilde{\Gamma}^{a\alpha\beta}\kappa_\beta (\tau)\; , \quad \nonumber \\  \delta_\kappa
e=-4i \kappa_\beta d\hat{\theta}^\beta \,
 \; ,  \qquad
  \end{eqnarray}
is infinitely reducible. It is parametrized by 32 component fermionic spinor function $\kappa_\beta (\tau)$ which however is not acting efficiently on the variable of the theory.\footnote{Roughly speaking,  due to the constraint $p_ap^a=0$,
$\kappa_\alpha$ and $\kappa_\alpha + p_a
\tilde{\Gamma}^{a}_{\alpha\beta}\kappa^{(1)\beta} (\tau)$ produce the same $\kappa$ variation of the Brink--Schwarz superparticle variables. One says that the above transformation has a null-vector $\kappa^{(1)\beta} (\tau)$ and, hence, the symmetry is {\it reducible}. But this is not the end of story.
One easily observes that $\kappa^{(1)\beta} (\tau)$ and $\kappa^{(1)\beta} (\tau)+p_a
\tilde{\Gamma}^{a\alpha\beta}\kappa^{(2)}_\beta (\tau)$, with an arbitrary $\kappa^{(2)}_\beta (\tau)$, makes the same change of the parameter $\kappa_\alpha$. This implies that there is a null--vector  for null--vector and that the $\kappa$--symmetry possesses at least the  second rank of reducibility. Furthermore, one sees that this process of finding higher null--vectors can be continued up to infinity (next stages are completely equivalent to the first two ones) so that one speaks about infinite reducibility of the $\kappa$--symmetry of the Brink--Schwarz superparticle. The number of the fermionic degrees of freedom which can be removed by $\kappa$--symmetry is then calculated as an infinite sum
$32-32+32-32+...= 32\cdot (1-1+1-1+...)= 32\cdot \lim\limits_{q\rightarrow 1} (1-q+q^2-...)= 32\cdot \lim\limits_{q\rightarrow 1}{1\over 1+q}= 16$. }

The irreducible $\kappa$--symmetry of the spinor moving frame formulation
(\ref{kap=irr}) can be obtained from the infinitely reducible (\ref{kappa}) by
substituting for $p_a$ the solution $p_a=\rho^{\#}u_a^=$ of the  constraint $p_ap^a=0$;
furthermore, using (\ref{Iu--=vGv}), we find
\begin{eqnarray}
\label{ep=kappa}
\epsilon^{+q}= 2
\rho^{\#}v_q^{-\alpha}\kappa_\alpha \; . \qquad
  \end{eqnarray}
Let us stress that this relation, as well as the transformation rules of the irreducible $\kappa$--symmetry (\ref{kap=irr}),  necessarily  involves the constrained spinors $v_q^{-\alpha}$. Thus the covariant irreducible  form of the $\kappa$--symmetry is a characteristic property of the spinor moving frame and similar ('twistor--like') formulations of the superparticle mechanics.\footnote{ Notice that in D=3,4 and 6 dimensions the counterpart of $v_q^{-\alpha}$ can be chosen to be unconstrained spinors; see references in {\it e.g.} \cite{B90,BL98',IB07:M0}.}

The importance of the $\kappa$--symmetry is related to the fact that it  reflects a
part of target space  supersymmetry which is preserved by ground state of the brane
under consideration \cite{Bergshoeff:1997kr,BdAI=2002,Bandos:2005ww} thus insuring that it is a  BPS
state. Its irreducible form, reached in the frame of spinor moving frame formulation, is useful not only for clarifying its nature as worldline supersymmetry  (\cite{stv}), but also for finding the corresponding induced supergravity multiplet which is necessary for constructing the mM$0$ action. To address this issue we need to comment on some properties of moving frame and spinor moving frame variables.

\subsection{Moving frame and spinor moving frame}
\label{smovfr}
To clarify  the origin and nature of the set of spinors $v^{-\alpha}_q$ which provide the square root of the light--like vector $u_a^{=}$ in the sense of Eqs. (\ref{Iu--=vGv}), and which have been used to present the $\kappa$--symmetry in the irreducible form (\ref{kap=irr}), it is useful to complete the light--like vector $u_a^{=}$ till the  {\it moving frame} matrix,
 \begin{eqnarray}\label{Uin}
& U_b^{(a)}= \left({u_b^{=}+ u_b^{\#}\over 2}, u_b^{i}, { u_b^{\#}-u_b^{=}\over 2}
\right)\; \in \; SO(1,10)\;  \quad
\end{eqnarray}
($i=1,...,9$). The statement that this matrix is an  element of the $SO(1,10)$, having been made in (\ref{Uin}),  is tantamount to saying that \begin{eqnarray}\label{UTetaU} U^T\eta U=I\; , \quad \eta^{ab}=diag (+1,-1,...,-1)\; , \quad
\end{eqnarray} which in its turn implies that the moving frame vectors obey the following set of constraints \cite{Sok}
\begin{eqnarray}\label{u--u--=0}
u_{ {a}}^{=} u^{ {a}\; =}=0\; , \quad    u_{ {a}}^{=} u^{ {a}\,i}=0\; , \qquad u_{
{a}}^{\; = } u^{ {a} \#}= 2\; , \qquad
 \\  \label{u++u++=0} u_{ {a}}^{\# } u^{ {a} \#
}=0 \; , \qquad
 u_{{a}}^{\;\#} u^{ {a} i}=0\; , \qquad  \\  \label{uiuj=-} u_{ {a}}^{ i}
 u^{{a}j}=-\delta^{ij}.  \quad
\end{eqnarray}
The 11D {\it spinor moving frame variables} (appropriate for our case) can be defined as $16\times 32$ blocks  of the $Spin(1,10)$ valued
matrix  \begin{eqnarray}\label{harmVin} V_{(\beta)}^{\;\;\; \alpha}=
\left(\begin{matrix}  v^{+\alpha}_q
 \cr  v^{-\alpha}_q \end{matrix} \right) \in Spin(1,10)\;
 \;  \qquad
\end{eqnarray}
double covering the moving frame matrix (\ref{Uin}). This statement implies that the similarity transformations with the matrix $V$ leave the 11D charge conjugation matrix invariant and, when applied to the 11D Dirac matrices, produce the same effect as 11D Lorentz rotation with   matrix $U$,
\begin{eqnarray}\label{VGVT=UG} \label{VCV=C}
VCV^T=C \; ,  \qquad \\
V\Gamma_b V^T =  U_b^{(a)} {\Gamma}_{(a)}\; , \qquad \\ \label{VTGV=UG} V^T
\tilde{\Gamma}^{(a)}  V = \tilde{\Gamma}^{b} u_b^{(a)}  \; . \qquad
\end{eqnarray} The two seemingly mysterious constraints (\ref{Iu--=vGv}) appear as a 16$\times$16 block of the second of these relations, (\ref{VGVT=UG}),
and as a component   $V^T \tilde{\Gamma}^{=}  V = \tilde{\Gamma}^{b} u_b^{=}$ of the third one, (\ref{VTGV=UG})  (with an appropriate representation of the 11D Gamma matrices). The other blocks/components of these constraints involve the second set of constrained spinors,
\begin{eqnarray}\label{M0:v+v+=u++}
 v_{q}^+ {\Gamma}_{ {a}} v_{p}^+ = \; u_{ {a}}^{\# } \delta_{qp}\; , \qquad
 v_{q}^- {\Gamma}_{ {a}} v_{p}^+ = - u_{ {a}}^{i} \gamma^i_{qp}\; , \qquad
\end{eqnarray}
\begin{eqnarray}\label{M0:u++G=v+v+}
 2 v_{q}^{+ {\alpha}}v_{q}^{+}{}^{ {\beta}}= \tilde{\Gamma}^{ {a} {\alpha} {\beta}} u_{
 {a}}^{\# }\; , \quad
 2 v_{q}^{-( {\alpha}}v_{q}^{+}{}^{ {\beta})}=-  \tilde{\Gamma}^{ {a} {\alpha} {\beta}}
 u_{ {a}}^{i}\; . \quad
\end{eqnarray}
Here $\gamma^i_{qp}$ are the 9d Dirac matrices; they are real, symmetric, $\gamma^i_{qp}=\gamma^i_{pq}$,  and obey the
Clifford algebra \begin{eqnarray}\label{gigj+=} \gamma^i\gamma^j + \gamma^j \gamma^i=
2\delta^{ij} I_{16\times 16}\; , \qquad
\end{eqnarray}
as well as the following identities
\begin{eqnarray}\label{gi=id1}
&& \gamma^{i}_{q(p_1}\gamma^{i}_{p_2p_3) }= \delta_{q(p_1}\delta_{p_2p_3) }\; , \qquad
\\ \label{gi=id2} && \gamma^{ij}_{q(q^\prime }\gamma^{i}_{p^\prime)p }+
\gamma^{ij}_{p(q^\prime }\gamma^{i}_{p^\prime)q } = \gamma^{j}_{q^\prime
p^\prime}\delta_{qp}-\delta_{q^\prime p^\prime}\gamma^{j}_{qp} \; .
\end{eqnarray}

Thus $ v_{q}^{- {\alpha}}$ and $ v_{q}^{+{\alpha}}$ can be identified as  square roots
of the light--like vectors $u_{ {a}}^{=}$ and $u_{ {a}}^{\# }$, respectively, while to
construct $u_{ {a}}^{i}$ one needs both these sets of constrained spinors.

The first constraint,  eq. (\ref{VCV=C}), implies that the inverse spinor moving frame matrix
\begin{eqnarray}\label{Vharm=M0}
 V^{( {\beta})}_{ {\alpha}}= \left(
v_{ {\alpha}q}{}^+\, ,v_{ {\alpha}q}{}^- \right)\; \in \; Spin(1,10) \; ,  \qquad  \\ \nonumber  V_{( {\beta})}{}^{ {\gamma}}
V_{ {\gamma}}^{ ({\alpha})}=\delta_{( {\beta})}{}^{ ({\alpha})}=\left(\begin{matrix}
\delta_{qp} & 0           \cr
          0 & \delta_{qp} \end{matrix}\right) \qquad  \\ \nonumber \Leftrightarrow  \quad \begin{cases} v_{q}^{- {\alpha}}v_{ {\alpha}p}{}^+=\delta_{qp}= v_{q}^{+
{\alpha}}v_{ {\alpha}p}{}^-\, , \cr  v_{q}^{- {\alpha}}v_{ {\alpha}p}{}^-= 0\; =
v_{q}^{+ {\alpha}}v_{ {\alpha}p}{}^+\, , \end{cases}\;
\end{eqnarray}
can be constructed from  $ v_{q}^{\mp {\alpha}}$,
\begin{eqnarray}
\label{V-1=CV}  v_{\alpha}{}^{-}_q =  i C_{\alpha\beta}v_{q}^{- \beta }\, ,
\qquad v_{\alpha}{}^{+}_q = - i C_{\alpha\beta}v_{q}^{+ \beta }\, .
 \end{eqnarray}

\subsection{Cartan forms, differentiation and variation of the (spinor) moving frame
variables }
 \label{DuDv}

To vary the action and to clarify the structure of the equations of motion one needs to vary  and to
differentiate the moving frame and spinor moving frame variables.  As these
are constrained, at the first glance this problem might look complicated, but,
actually, this is not the case. The clear group theoretical structure beyond the moving
frame and spinor moving frame variables makes their differential calculus and
variational problem extremely simple.

Referring again for the details to  \cite{IB07:M0,mM0=PLB}, let us just state that the
derivatives of the moving frame and spinor moving frame variables can be expressed in
terms of the ${so(1,10)}$--valued Cartan forms $\Omega^{(a)(b)}= U^{(a)c}dU_c^{(b)}$
the set of which can be split onto the covariant Cartan forms
\begin{eqnarray}
\label{Om++i=} \Omega^{=i}= u^{=a}du_a^{i}\; , \qquad \Omega^{\# i}=  u^{\#
a}du_a^{i}\; , \qquad
  \end{eqnarray}
providing the basis for the coset ${SO(1,10)\over SO(1,1)\times SO(9)}$, and the forms
\begin{eqnarray}
\label{Om0:=} \Omega^{(0)}= {1\over 4} u^{=a}du_a^{\#}\; , \qquad \\ \label{Omij:=}
\Omega^{ij}=  u^{ia}du_a^{j}\; , \qquad
  \end{eqnarray}
which have the properties of the $SO(1,1)$ and $SO(9)$ connection respectively. These
can be used to define the $SO(1,1)\times SO(9)$ covariant derivative $D$. The covariant
derivative of the moving frame vectors is expressed in terms of the  covariant Cartan
forms (\ref{Om++i=})
\begin{eqnarray}\label{M0:Du--=Om}
Du_{ {b}}{}^{=} &:= & du_{ {b}}{}^{=} +2 \Omega^{(0)} u_{ {b}}{}^{=}= u_{ {b}}{}^i
\Omega^{= i}\; , \qquad \\ \label{M0:Du++=Om} Du_{ {b}}{}^{\#}&:=& du_{ {b}}{}^{\#} -2
\Omega^{(0)} u_{ {b}}{}^{\#}=  u_{ {b}}{}^i \Omega^{\# i}\; , \qquad \\
\label{M0:Dui=Om}  Du_{ {b}}{}^i &:=& du_{ {b}}{}^{i} - \Omega^{ij} u_{ {b}}{}^{j} =
{1\over 2} u_{ {b}}{}^{\# } \Omega^{=i}+ {1\over 2}  u_{ {b}}{}^{=} \Omega^{\#
i} . \; \qquad
\end{eqnarray}
The same is true for the spinor moving frame variables,
\begin{eqnarray}
\label{Dv-q}  Dv_q^{-\alpha}&:=& dv_q^{-\alpha} +  \Omega^{(0)} v_q^{-\alpha} - {1\over
4}\Omega^{ij} \gamma^{ij}_{qp} v_p^{-\alpha} = \nonumber \\ &=& - {1\over 2}
\Omega^{=i} v_p^{+\alpha} \gamma_{pq}^{i}\; , \qquad \\ \label{Dv+q}  Dv_q^{+\alpha}
&:=& dv_q^{+\alpha} -  \Omega^{(0)} v_q^{+\alpha} - {1\over 4}\Omega^{ij}
\gamma^{ij}_{qp} v_p^{+\alpha} = \nonumber \\ &=& - {1\over 2} \Omega^{\# i}
v_p^{-\alpha} \gamma_{pq}^{i}\; . \qquad
\end{eqnarray}

The variation of moving frame and spinor moving frame variables can be obtained from
the above expression for derivatives by a formal contraction with variation symbol,
$i_\delta d=\delta $ (this is to say, by taking the Lie derivatives). The independent
variations are then described by $i_\delta$ contraction of the Cartan forms, $i_\delta
\Omega^{(a)(b)}$. Furthermore, $i_\delta \Omega^{(0)}$ and $i_\delta \Omega^{ij}$ are
the parameters of the $SO(1,1)$ and $SO(9)$ transformations, which are manifest gauge
symmetries of the model. Then the essential variation of the moving frame and spinor
moving frame variables, this is to say, variations which produce (better to say, which
may produce) nontrivial equations of motion, are expressed in terms of $i_\delta
\Omega^{=i}$ and $i_\delta \Omega^{\# i}$,
\begin{eqnarray}\label{vu--=iOm}
\delta u_{ {b}}{}^{=} = u_{ {b}}{}^i i_\delta\Omega^{= i}\; , \qquad \label{vu++=iOm}
\delta u_{ {b}}{}^{\#}=  u_{ {b}}{}^i i_\delta\Omega^{\# i}\; , \qquad \\
\label{vui=iOm}  \delta u_{ {b}}{}^i  = {1\over 2} \, u_{ {b}}{}^{\# }  i_\delta
\Omega^{=i}+ {1\over 2} \, u_{ {b}}{}^{=}  i_\delta\Omega^{\# i}\; . \qquad
\\
\label{vv-q}  \delta v_q^{-\alpha}= - {1\over 2} i_\delta \Omega^{=i} v_p^{+\alpha}
\gamma_{pq}^{i}\; , \qquad \\ \label{vv+q}  \delta v_q^{+\alpha} = - {1\over 2}
i_\delta\Omega^{\# i} v_p^{-\alpha} \gamma_{pq}^{i}\; . \qquad
\end{eqnarray}

\subsection{$K_9$ gauge symmetry of the spinor moving frame action of the M$0$-brane}
\label{secK9}
A simple application of the above formulae begins by
observing that the parameter $i_\delta\Omega^{\# i}$ does not enter the variation of neither $u_a^=$ nor $v^{-\alpha}_q$. However, the  M$0$-brane (\ref{SM0=}), (\ref{SM0==}) involves only
these (spinor) moving frame variables. Hence the transformation of the spinor moving frame corresponding to $\tau$ dependent parameters $k^{\# i}=i_\delta\Omega^{\# i}$ are gauge symmetries
of this M$0$  action. These so--called $K_9$--symmetry transformations
\begin{eqnarray}\label{vK9}
\delta u_{ {b}}{}^{=} = 0  , \quad
\delta u_{ {b}}{}^{\#}=  u_{ {b}}{}^i k^{\# i}, \quad \delta u_{ {b}}{}^i  =  {1\over 2} \, u_{ {b}}{}^{=}  k^{\# i}\; , \quad
\\
\label{vK9v}  \delta v_q^{-\alpha}= 0 , \qquad \delta v_q^{+\alpha} = - {1\over 2}
k^{\# i} v_p^{-\alpha} \gamma_{pq}^{i}\;  \qquad
\end{eqnarray}
should be taken into account when calculating the number of M$0$ degrees of freedom.

Quite interesting remnant of this K9 symmetry survives in the multiple M$0$ case and will be essential to understand the  structure of mM$0$ equations of motion.

\subsection{Derivatives and variations of the Cartan forms}
\label{secDOm}

One can easily check that the covariant Cartan forms are covariantly constant,
\begin{eqnarray}\label{M0:DOm--=} D\Omega^{= i}=  0\; , \qquad D\Omega^{\# i}  = 0\;  ,  \qquad
\end{eqnarray}
where the covariant derivatives include the induced connection (\ref{Om0:=}), (\ref{Omij:=}) \footnote{ $D\Omega^{= i}:=d\Omega^{= i} +2 \Omega^{= i}\wedge \Omega^{(0)}+ \Omega^{= j}\wedge \Omega^{ji}$, see  (\ref{M0:Du--=Om})--(\ref{M0:Dui=Om}).}.   The curvatures of these connections are
\begin{eqnarray}\label{M0:Gauss}
 && F^{(0)}:= d\Omega^{(0)} =    {1\over 4 } \Omega^{=\, i} \wedge
 \Omega^{\# \, i}\; , \qquad  \\
\label{M0:Ricci} && {G}^{ij}:= d\Omega^{ij}+ \Omega^{ik} \wedge \Omega^{kj} = - \Omega^{=\,[i} \wedge \Omega^{\# \, j]}\;
,  \qquad
\end{eqnarray}
can be calculated, e.g., from the integrability conditions of Eqs. (\ref{M0:Du--=Om})--(\ref{M0:Dui=Om}),
 \begin{eqnarray}
\label{DDu++=} && DDu_{ {a}}^{\# } = - 2 F^{(0)}u_{ {a}}^{\#
}\; , \qquad
 DDu_{ {a}}{}^{i} = u_{ {a}}^{j} {G}^{ji}  \; . \qquad
\end{eqnarray}
As in the case of moving frame variables (see sec. \ref{DuDv}), the variations of the Cartan forms can be obtained  from the above expressions using the Lie derivative formula. Omitting the transformations of manifest gauge symmetries SO(1,1) and SO(9) (parametrized by $i_\delta \Omega^{(0)}$ and $i_\delta \Omega^{ij}$), we present the essential variations:
\begin{eqnarray}
\label{vOm++=} \delta \Omega^{\# i}&=& D i_\delta \Omega^{\# i}\; , \qquad  \delta \Omega^{=i}= D i_\delta \Omega^{=i}\; , \qquad
 \\
\label{vOmij=} \delta \Omega^{ij}\; &=& - \Omega^{=[i} i_\delta \Omega^{\# j]} -  \Omega^{\# [i}i_\delta \Omega^{=j]}\; , \qquad
\\
\label{vOm0=}
\delta \Omega^{(0)}&=& \frac {1}{4} \Omega^{=i}i_\delta \Omega^{\# i} -
 \frac {1}{4}  \Omega^{\# i}i_\delta \Omega^{=i}  \; . \qquad
\end{eqnarray}
These equations will be useful  to vary the multiple M$0$ brane action in Sec. \ref{SecEqmM0}. For deriving the equations of motion of single M$0$ brane it is sufficient to use Eqs. (\ref{vu--=iOm}), (\ref{vv-q}) and (\ref{M0:Du--=Om})--(\ref{Dv+q}).

\section{Equations of motion of a single M$0$ brane and induced ${\cal N}=16$ supergravity on the worldline $W^1$}
\label{M0eqs}
\setcounter{equation}{0}

In this section we summarize the equation of motion for a single M$0$-brane obtained from the spinor moving frame action \cite{IB07:M0} and discuss the
$1d$ ${\cal N}=16$ supergravity multiplet induced by embedding of the worldline in target 11D superspace. This induced supergravity will be used to write the mM$0$ action.  We also show here that a supersymmetric solutions of the equations of motion of single M$0$-brane preserve just one half of the 11D supersymmetry. This result, seemingly new although not unexpected, is necessary to make similar conclusion about supersymmetric solutions of the mM$0$ equations.

\subsection{Equations of motion for spinor moving frame variables}

The moving frame matrix $U_a^{(b)}$ (\ref{Uin}) provides a `bridge' between the 11D Lorentz group and its $SO(9)\otimes SO(1,1)$ subgroup in the sense that it carries one index ($_a$) of $SO(1,10)$ and one index ($^{(b)}$) transformed by a matrix from $SO(9)\otimes SO(1,1)$ subgroup of $SO(1,10)$. Contracting the pull--back of the bosonic supervielbein form $\hat{E}^b$ we arrive at
 \begin{eqnarray}\label{EU=EEE}
\hat{E}^{(a)}=\hat{E}^b U_b^{(a)}= ( \hat{E}^{=},
\hat{E}^{\#}, \hat{E}^i)
\;   \qquad
\end{eqnarray}
which is split covariantly in three types of one forms. These are inert under $SO(1,10)$ but carry the nontrivial SO(9) vector index (in the case of  $\hat{E}^i$) or $SO(1,1)$ weights (in the cases of $\hat{E}^{=}$ and $
\hat{E}^{\#}$). The corresponding decomposition of the vector representation of  $SO(1,10)$ with respect to
its $SO(9)\otimes SO(1,1)$ subgroup,
$$ {\bf 11}\mapsto {\bf 1}_{-2}+  {\bf 1}_{+2}+  {\bf 9}_{0}\; , $$
is even better illustrated by the equation $\hat{E}^{(a)}U_{(a)}{}^b=\hat{E}^b $ which, in more detail, reads
\begin{eqnarray}\label{E=E+E+E}
\hat{E}^{a}= {1\over 2} \hat{E}^{=} u^{a\#}+
{1\over 2}\hat{E}^{\#}u^{a=} - \hat{E}^i u^{ai}
\; .  \qquad
\end{eqnarray}
 Thus the  moving frame vectors help to split the pull--back of the supervielbein in a
Lorentz covariant manner. The $SO(9)$ singlet one form with $SO(1,1)$ weight -2,  $\hat{E}^{=}=
\hat{E}^b u_b^{=}$ enters the action (\ref{SM0=}) multiplied by the weight +2 worldline scalar field $\rho^{\#}(\tau)$. This clearly has the meaning of the Lagrange multiplier: its variation results in vanishing of $\hat{E}^{=}$,
\begin{eqnarray}\label{E==0}
\hat{E}^{=}:= \hat{E}^a u_{ {a}}^{=}=0\; .  \qquad
\end{eqnarray}
Now, the variation of $\hat{E}^{=}$ contains two different contributions, $\delta \hat{E}^{=}= \delta\hat{E}^a u_a^{=}+ \hat{E}^a \delta u_a^{=}$. The first comes from the variation of the  pull--back of the bosonic supervielbein form which in our case of flat target superspace can be easily calculated with the result
\begin{eqnarray}\label{vhEa=}
\delta\hat{E}^a = - i d\hat{\theta}\Gamma^a\delta\hat{\theta}+ d(\delta \hat{x}^a -i \delta\hat{\theta} \Gamma^a\hat{\theta}) \; .  \qquad
\end{eqnarray}
The second term contains the variation of the light--like vector $u_a^{=}$ which can be written as in Eq. (\ref{vu--=iOm}), $\delta u_a^{=}= u_a^{i}\, i_\delta \Omega^{=i}$ with an arbitrary $i_\delta \Omega^{=i}$. The corresponding variation of the action (\ref{SM0=}) reads  $\delta_u S_{M0}= \int_{W^1} \rho^{\#}\,
\delta u_a^{=}  \, \hat{E}^a = \int_{W^1} \rho^{\#}\,
u_a^{i}  \, \hat{E}^a i_\delta \Omega^{=i} $ and produce the equation of motion
\begin{eqnarray}\label{Ei=0}
\hat{E}^{i}:=
\hat{E}^a u_{ {a}}^{i}=0\;  .  \qquad
\end{eqnarray}
 Using Eq. (\ref{E=E+E+E}) one can collect Eqs. (\ref{E==0}) and (\ref{Ei=0}) in
\begin{eqnarray}\label{E==0=Ei}
\hat{E}^{a}:= {1\over 2}\hat{E}^{\#} u^{a=} \; .  \qquad
\end{eqnarray}
This equation shows that the M$0$--brane worldline $W^1$ is a light--like line in target
(super)space, as it should be for the massless superparticle.

\subsection{Induced supergravity on the worldline of single M$0$-brane}

Furthermore (\ref{E==0=Ei}) suggests to consider $\hat{E}^{\#}$ as einbein on the worldline $W^1$; this composite einbein is induced by embedding of $W^1$ into the target superspace. The transformation of  $\hat{E}^{\#}$ under the irreducible  $\kappa$--symmetry (\ref{kap=irr}) is given by
$\delta_\kappa\hat{E}^{\#}= -2i \hat{E}^{+q}\epsilon^{+q}$. In the light of the identification of $\kappa$--symmetry with local worldline supersymmetry \cite{stv}, this equation suggests
to consider the covariant {\bf 16}$_{+}$ projection, $ \hat{E}^{+q}=
\hat{E}^{\alpha}v_\alpha^{+q}$, of the pull--back of the fermionic 1--form $E^\alpha$ as induced `gravitino' companion of the induced 1d `graviton'  $\hat{E}^{\#}$. Indeed under the $\kappa$--symmetry  (\ref{kap=irr}) this set of forms  shows the typical transformations rules of
(1d ${\cal N}=16$) supergravity multiplet,
\begin{eqnarray}\label{v1dSG=}
\delta_\kappa \hat{E}^{+q}= D \epsilon^{+q}(\tau) \; ,  \qquad \delta_\kappa
\hat{E}^{\#}= -2i \hat{E}^{+q}\epsilon^{+q}\; .
\end{eqnarray}
Here $D=d\tau D_\tau$ is the $SO(1,1)\times SO(9)$ covariant derivative which we will
specify below. The connections in this covariant derivative are defined in terms of moving frame variables and, hence, are inert under the $\kappa$--symmetry; in this sense the induced 1d ${\cal N}=16$ supergravity multiplet is described essentially by 1 bosonic and 16 fermionic 1--forms $\hat{E}^{\#}$ and $\hat{E}^{+q}$.
Our  action for the mM$0$ system, which we present in the next section,
will contain the coupling of this induced 1d supergravity to the matter describing the
relative motion of the mM$0$ constituents.

\subsection{Dynamical equations of single M$0$-brane}

The other, {\bf 16}$_-$ projection $\hat{E}^{-q}=  \hat{E}^{\alpha}v_\alpha^{-q}$ of the pull--back of fermionic supervielbein form to $W^1$  vanishes on the mass shell,
\begin{eqnarray}\label{E-q=0}
\hat{E}^{-q}:=  \hat{E}^{\alpha}v_\alpha^{-q}=0  \; .  \qquad
\end{eqnarray}
Indeed, varying the coordinate functions in the action (\ref{SM0=}) we arrive at equation ${\delta S_{M0}\over \delta \hat{Z}^M}=0$  which reads
\begin{eqnarray}
\label{vhZ=>}\partial_\tau(\rho^{\#} u_a^{=}E_M^a(\hat{Z}))=0\; .
\end{eqnarray}
In our case of flat target superspace
$E_M^a(\hat{Z})= \delta_M^a -i \delta_M^\alpha (\Gamma^a\hat{\theta})_\alpha $ and one can easily split  (\ref{vhZ=>}) into the bosonic vector and fermionic spinor equations (which we prefer to write with the use of $d=d\tau \partial_\tau$)
\begin{eqnarray}
\label{vhx=>} d( \rho^{\#} u_a^{=})=0\; , \\
\label{vhth=>} \rho^{\#} u_a^{=}( \Gamma^a\partial_\tau\hat{\theta} )_\alpha =0\; .
\end{eqnarray}

Using (\ref{Iu--=vGv}b) and assuming $\rho^{\#}\not= 0$ we find that (\ref{vhth=>}) is equivalent to  Eq. (\ref{E-q=0}). This implies that the $d\hat{\theta}^\alpha$ can be expressed through the induced gravitino,
\begin{eqnarray}\label{Ef=E+v-}
\hat{E}^{\alpha}=d\hat{\theta}^\alpha = \hat{E}^{+q} v_q^{-\alpha} \; .  \qquad
\end{eqnarray}

Let us come back to the equation for the bosonic coordinate functions, (\ref{vhx=>}) (or equivalently, $\partial_\tau( \rho^{\#} u_a^{=})=0$).
Using  (\ref{M0:Du--=Om}) we can write this in the form $0= D\rho^{\#} \, u_a^{=} +
\rho^{\#}  u_a^i\Omega^{=i}$. Here and below we use the covariant derivatives defined in (\ref{M0:Du--=Om}), (\ref{M0:Du++=Om}), (\ref{M0:Dui=Om})). Contracting that equation with  $ u^{a\#}$ gives us
\begin{eqnarray}\label{M0:Drho=0}
 D\rho^{\#}=0 \; ,
\end{eqnarray}
while  the nontrivial part of the bosonic equations of motion of a single M$0$--brane,
which can be read off from the coefficient for $u_a^i$, states that the covariant Cartan
form $\Omega^{= i}$ vanishes,
\begin{eqnarray}
\label{Om--i=0} \Omega^{= i}=0 \; . \qquad
\end{eqnarray}
Coming back to Eq.  (\ref{M0:Du--=Om}), we see that Eq. (\ref{Om--i=0}) can be
expressed by stating that the covariant derivative of the light--like vector $u_a^=$
vanishes,
\begin{eqnarray}
\label{Du--=0} Du_a^{=}=0 \; ,  \qquad
\end{eqnarray}
or, equivalently, by
\begin{eqnarray}
\label{Dv-q=0} Dv_q^{-\alpha}=0 \; .  \qquad
\end{eqnarray}

On the other hand, using \begin{eqnarray}
\label{D++E++i=0}
D=d\tau D_\tau= \hat{E}^\# D_\# \; , \qquad
\end{eqnarray}  we can write Eq. (\ref{Du--=0}) in the
form $D_\# u_a^{=}=0$, and, as far as (\ref{E==0=Ei}) implies $u_a^{=}=2 \hat{E}_\#^a$,
in the following more standard form
\begin{eqnarray}
\label{D++E++i=0} D_\# \hat{E}_\#^a =0 \; ,  \qquad
\end{eqnarray}
or, in more detail,
\begin{eqnarray}
\label{D++D++x=0} D_\# D_\# \hat{x}^a=  i D_\#(D_\# \hat{\theta}\Gamma^a\hat{\theta})\;
.  \qquad
\end{eqnarray}

Two more observations will be useful below. The first is that Eq. (\ref{M0:Drho=0}), $0=D\rho^{\#}=d\rho^{\#}-2\rho^{\#}\Omega^{(0)}$, can be solved with respect to the induced $SO(1,1)$ connection,
\begin{eqnarray}
\label{M0:Om0=}
\Omega^{(0)} = {d\rho^{\#}\over 2\rho^{\#}} \; .  \qquad
\end{eqnarray}
Notice that this is in agreement with the statement that one can always gauge away any 1d connection: using the local SO(1,1) symmetry to fix the gauge $\rho^{\#}=const$ we arrive at $\Omega^{(0)} =0$.

The second comment concerns the supersymmetric pure bosonic solutions of the above equations of motion.

\subsection{All supersymmetric solutions of the M$0$ equations describe 1/2 BPS
states}
\label{susySOL}

As far as the fermionic coordinate function $\hat{\theta}^\alpha$ is transformed by both spacetime supersymmetry and by the  worldline supersymmetry ($\kappa$--symmetry) (\ref{susy-th}), $\delta \hat{\theta}^\alpha=
-{\varepsilon}^\alpha + \epsilon^{+q}(\tau) v^{-\alpha}_q(\tau)$,
the purely bosonic solutions of the M$0$ equations, having
\begin{eqnarray}\label{hth=0}
 \hat{\theta}^\alpha = 0 \; ,
\end{eqnarray}
may preserve  a part of target space supersymmetry. This is characterized by parameter
\begin{eqnarray}\label{vep=epv}
 {\varepsilon}^\alpha = \epsilon^{+q}(\tau) v^{-\alpha}_q(\tau) \; .
\end{eqnarray}
The left hand side of this equation contains a constant fermionic spinor
$d{\varepsilon}^\alpha=0$, so that $d(\epsilon^{+q} v^{-\alpha}_q)=D\epsilon^{+q}
v^{-\alpha}_q+ \epsilon^{+q} Dv^{-\alpha}_q=0$. Furthermore, taking into account that
the equations of motion for the bosonic coordinate function, Eq.  (\ref{D++D++x=0}),
implies  (\ref{Dv-q=0}), one finds that the consistency of (\ref{vep=epv}) is the
covariant constancy of the $\kappa$--symmetry parameter $\epsilon^{+q}(\tau)$,
\begin{eqnarray}\label{Dep+q=0}
 D\epsilon^{+q}=0
\; .
\end{eqnarray}
In 1d system the connection can be gauged away so that this condition can be reduced to the
existence of a constant $SO(9)$ spinor  $\epsilon^q$. For instance gauging away the
$SO(9)$ connection and using Eq. (\ref{M0:Om0=}), we can present (\ref{Dep+q=0}) in
the form  $d(\epsilon^{+q}/\sqrt{\rho^{\#}})=0$ and solve it by $\epsilon^{+q}=
 \sqrt{\rho^{\#}}\, \epsilon^q$ with $d\epsilon^q=0$.

This implies that {\it any purely bosonic solution of the M$0$ equations preserves
exactly $1/2$ of the spacetime supersymmetry}.

\section{Covariant action for multiple M0--brane system}
\label{mM0secAC}
\setcounter{equation}{0}

In this section we obtain  the action for mM$0$ system, first presented in  \cite{mM0=action}. We start with $1d$ ${\cal N}=16$ $SU(N)$ SYM action, make it supersymmetry local by coupling to $1d$ ${\cal N}=16$ supergravity, and add to such a locally supersymmetric functional  the counterpart of a single M$0$ action for the  center of energy variables which induces the above supergravity multiplet on the center of energy worldline. The local supersymmetry of the induced supergravity is produced by a generalization of the $\kappa$--symmetry transformations of single M$0$--brane acting on the center of energy variables.

\subsection{Variables describing the mM0 system}

Let us introduce the dynamical variables describing the system of multiple M$0$ branes, which we abbreviate as mM$0$. Its dimensional reduction  is expected to produce the system of N nearly coincident D$0$-branes (mD$0$ system) and at very low energy this later is described by the action of 1d ${\cal N}=16$ supersymmetric Yang--Mills theory (SYM) with the gauge group $U(N)$, which is given by dimensional reduction of the 10D ${\cal N}=1$ U(N) SYM down to d=1. Now,  the set of fields of the U(N) SYM  can be split onto the non-Abelian SU(N) SYM and Abelian U(1) SYM multiplets. Roughly speaking, this later describes the center of energy motion of the mD$0$ system while the former corresponds to the relative motion of the constituents of the mD$0$ system. Then it is natural to assume that the relative motion of the mM$0$ constituents is also described by the fields of $SU(N)$ SYM multiplet.

Now let us turn to the center of energy motion. We begin by noticing that the $U(1)$ SYM  fields can be seen in the single D$0$ brane action (see \cite{B+T=Dpac} and refs therein) after fixing the gauge with respect to $\kappa$--symmetry and reparametrization symmetry. Originally the action of a single D$0$ brane is written in terms of 10 bosonic and $32$ fermionic coordinate functions, worldline fields  corresponding to the coordinates of type IIA $D=10$ superspace. The above gauge fixing reduces the number of fermionic fields to $16$ and the number of bosonic coordinate functions to $9$. These are the same as the number of physical fields as 1d reduction of the 10D SYM theory. This also contains the time component of the gauge field which can be gauged away by the U(1) gauge symmetry transformation  and do not carry degrees of freedom. The U(1) SYM multiplet describing the center of energy motion of the mD$0$ system can be obtained  by fixing the gauge with respect to  $\kappa$-symmetry and reparametrization symmetry on the coordinate functions, the same as in the case of single D$0$ brane.

In the light of the above discussion, it is natural to describe the center of energy motion of the mM$0$ system by the 11 bosonic and 32 fermionic  coordinate functions, the same  as used to describe the motion of single M$0$ brane, and to assume that the wanted  mM$0$  action possesses $\kappa$--symmetry and reparametrization symmetry, like the single M$0$-brane action does.

To resume, following \cite{mM0=PLB,mM0=PRL,mM0=action} we will describe the center of energy motion of $N$ nearly coincident M$0$-branes (mM$0$ system) by the 11 commuting and 32 anti-commuting coordinate functions
\begin{eqnarray}\label{hZ=hx+}
\hat{Z}{}^M(\tau)&=& (\hat{x}{}^\mu (\tau), \hat{\theta}{}^\alpha (\tau))\; , \quad \\ \nonumber && \mu =0,1,...,10; \quad
\alpha = 1,2,..., 32
\end{eqnarray}
(the same as used to describe single M$0$--brane), and the  relative motion of the mM$0$ constituents by the fields of the $SU(N)$ SYM supermultiplet. These are the bosonic and fermionic hermitian traceless $N\times N$ matrices
fields   \begin{eqnarray}\label{NxNXi}
{\bb X}^i(\tau)\quad & and & \quad   \Psi_q (\tau)
\\ \nonumber && (i=1,...,9\, , \qquad q=1,...,16)
\end{eqnarray}
 depending on a (center of energy)
proper time variable $\tau$. The bosonic ${\bb  X}^i(\tau)$ carries the index
$i=1,...,9$ of the vector representation of $SO(9)$, while the fermionic $\Psi_q$
transforms as a spinor under $SO(9)$,  $q=1,...,16$.

\subsection{First order form of the 1d ${\cal N}=16$ SYM Lagrangian as a starting point to build mM$0$ action}

The standard 1d ${\cal N}=16$ SYM Lagrangian (obtained by dimensional reduction of 10D SYM) can be written in the following first order form
\begin{eqnarray}
\label{LSYM=1}
d\tau L_{SYM}= tr\left(- {\bb P}^i \nabla_\tau {\bb X}^i + 4i { \Psi}_q \nabla_\tau
{\Psi}_q  \right) +   d\tau {\cal H}  \;   \qquad
  \end{eqnarray}
where the Hamiltonian
\begin{eqnarray}
\label{HSYM=1}
{\cal H}=   {1\over 2} tr\left( {\bb P}^i {\bb P}^i \right) + {\cal V} ({\bb X}) - 2\,  tr\left({\bb X}^i\, \Psi\gamma^i {\Psi}\right) \;   \qquad
  \end{eqnarray}
contains the positively definite scalar potential
\begin{eqnarray}
\label{VSYM=} {\cal V} = - {1\over 64}
tr\left[ {\bb X}^i ,{\bb X}^j \right]^2 \equiv  +{1\over 64}  tr\left[ {\bb X}^i
,{\bb X}^j \right] \cdot \left[ {\bb X}^i ,{\bb X}^j \right]^\dagger  . \quad
  \end{eqnarray}
Eqs. (\ref{LSYM=1}) involve the  auxiliary `momentum' fields, the nanoplet of traceless $N\times N$ matrices ${\bb P}^i$, and also the gauge field ${\bb A}_\tau (\tau)$ which enters the  covariant derivatives $\nabla=d\tau \nabla_\tau $ of the above bosonic and fermionic fields,
\begin{eqnarray}
\label{SYMDX=}
 \nabla {\bb  X}^i= d{\bb X}^i+ [{\bb A},    {\bb X}^i]\; , \qquad \nabla {\Psi}_q= d{\Psi}_q+ [{\bb A},   {\Psi}_q]\; .
  \end{eqnarray}
The action with the above Lagrangian is invariant under the following
d=1 ${\cal N}=16$  supersymmetry transformations with constant fermionic parameter $\varepsilon^q$
\begin{eqnarray}
\label{SYMsusy-X} \delta_\varepsilon {\bb X}^i   = 4i \varepsilon^q (\gamma^i  \Psi)_q \; , \quad
\delta_\varepsilon {\bb P}^i   = [\varepsilon^{q} (\gamma^i  \Psi)_q,  {\bb X}^j]\; ,\qquad \\
\label{SYMsusy-Psi} \delta_\varepsilon \Psi_q =  {1\over 2} \varepsilon^{p} \gamma^i_{pq}  {\bb
P}^i-  {i\over 16} \epsilon^{p} \gamma^{ij}_{pq}  [{\bb X}^i, {\bb X}^j]\; ,\qquad \\
\label{SYMsusy-A}
 \delta_\varepsilon {\bb A}  = - d\tau   \varepsilon^{q}  \Psi_q
 \; .  \qquad
\end{eqnarray}

The mM$0$ action should describe the coupling of the above SYM theory to the center of energy variables (\ref{hZ=hx+}). As we have discussed above, such an action should possess  the reparametrization symmetry and a 16 parametric local fermionic symmetry, a counterpart of the irreducible $\kappa$--symmetry (\ref{kap=irr}) of the single M$0$ action. It is natural also to think on this fermionic gauge symmetry as on the local version of the above rigid d=1 ${\cal N}=16$  supersymmetry of the SYM action,  Eqs. (\ref{SYMsusy-X})--(\ref{SYMsusy-A}).

\subsection{Induced supergravity on the center of energy worldline}

The natural way to make a supersymmetry local is to couple it to supergravity multiplet. As a by--product such a coupling should guaranty the reparametrization (general coordinate) invariance. Now it is the time to recall about induced supergravity multiplet on the worldline of the single M$0$ brane constructed in sec. (\ref{M0eqs}).
Similarly, we can associate a moving frame (\ref{Uin}) and spinor moving frame (\ref{harmVin}) to the center of energy motion of the mM$0$ system and use these together with center of energy coordinate functions (\ref{hZ=hx+}) to build the composite d=1 ${\cal N}=16$ supergravity multiplet including the 1d `graviton' and `gravitino'
\begin{eqnarray}
\label{E++=Eu++}
\hat{E}^{\#}&=& \hat{E}^{a}u_a^{\#}= (d\hat{x}^{a}- id\hat{\theta}\Gamma^a\hat{\theta}) u_a^{\#}\; , \qquad
\\ \label{E+q=Ev+q} \hat{E}^{+q}&=& \hat{E}^\alpha v_\alpha^{+q}=d\hat{\theta}^\alpha v_\alpha^{+q}
\; , \qquad
  \end{eqnarray}
transforming under the local supersymmetry as in (\ref{v1dSG=}),
\begin{eqnarray}\label{vcSG=}
\delta_\epsilon \hat{E}^{+q}= D \epsilon^{+q}(\tau) \; ,  \qquad \delta_\epsilon
\hat{E}^{\#}= -2i \hat{E}^{+q}\epsilon^{+q}\; .
\end{eqnarray}

Notice that the use of such a composite supergravity induced by  embedding of the center of energy worldline into the flat target 11D superspace implies  that the local supersymmetry parameter carries the weight +1
of the $SO(1,1)$ group transformations defined on the moving frame variables. This implies the necessity to adjust the $SO(1,1)$ weight also to the fields describing the relative motion of the mM$0$ constituents. Following \cite{mM0=PLB,mM0=PRL,mM0=action} we define the  $SO(1,1)$ weight of the bosonic and fermionic fields to be
 -2 and -3, respectively, so that in a more explicit notation (and using the conventions were the upper $^- $  index indicate the same -1 weight as the lower $_+ $ one)
 \footnote{Such a chose of weight of the basic matrix fields is preferable for the
description in the frame of superembedding approach, like developed in \cite{mM0=PLB,mM0=PRL}. Once using the density $\rho^{\#}=\rho^{++}$ which enters the spinor moving frame action for single  M$0$, we can easily change the weights of the fields  multiplying them  by corresponding power of  $\rho^{\#}$. However we find more convenient  to work with the `weighted' fields (\ref{bXi=bXi++}), (\ref{Psi=Psi+++}). }
\begin{eqnarray}
\label{bXi=bXi++} && {\bb  X}^i= {\bb  X}_{\#}^i:= {\bb  X}_{++}^i\, ,\qquad i=1,...,9\, , \qquad \\
\label{Psi=Psi+++} && {{ \Psi}}_q= \Psi_{\# \,+q}:= \Psi_{++ \,+q}= \Psi_{\#}{}_q^-\,  ,\quad q=1,...,16\, .
\qquad
 \end{eqnarray}

As in the case of single M$0$--brane, we expect the $SO(1,1)$ as well as $SO(9)$ transforation to be a gauge symmetry of our action. This implies the use of covariant derivative with   $SO(1,1)$ and $SO(9)$
connection. As in the case of single M$0$-brane, we define these connections to be constructed from the moving frame variables
\begin{eqnarray}
\label{mM0:Om0=} \Omega^{(0)}= {1\over 4} u^{=a}du_a^{\#}\; , \qquad \Omega^{ij}=
u^{ia}du_a^{j}\;  \qquad
  \end{eqnarray}
(see Eqs.  (\ref{Om0:=}) and (\ref{Omij:=})), which are now associated to the  center of energy motion of the mM$0$
system. The covariant derivatives of the su(N) valued matrix fields (\ref{bXi=bXi++}) are defined by
\begin{eqnarray}
\label{DXi=} D{\bb X}^i  &:=& d{\bb X}^i  + 2\Omega^{(0)} {\bb X}^i  - \Omega^{ij} {\bb
X}^j+ [{\bb A},    {\bb X}^i] \; , \qquad \\ \label{DPsi:=} D\Psi_q  &:=& d\Psi_q
 + 3\Omega^{(0)} \Psi_q   -{1\over 4} \Omega^{ij} \gamma^{ij}_{qp} {\Psi}_p+ [{\bb A},
 \Psi_q ] \; . \qquad
  \end{eqnarray}
They also involve the $SU(N)$ connection  ${\bb A}= d\tau {\bb A}_\tau (\tau)$  on the center of
energy worldline $W^1$. The anti-Hermitian traceless  $N\times N$ matrix gauge field ${\bb A}_\tau (\tau)$ is an independent variable of our model. Let us stress, however, that, as any 1d gauge field, it can be gauged away and thus does not carry any degree of freedom.

The covariant derivative of the supersymmetry parameter in (\ref{vcSG=}) reads
\begin{eqnarray}
\label{Dep+q=}
D\epsilon^{+q}= d\epsilon^{+q}- \Omega^{(0)} \epsilon^{+q}  +{1\over 4} \Omega^{ij} \epsilon^{+p}  \gamma^{ij}_{pq} \; , \qquad
\end{eqnarray}
so that the induced connection (\ref{mM0:Om0=}) are also the members of the composite d=1 ${\cal N}=16$ supergravity multiplet.

\subsection{A way towards mM$0$ action }

Now we are ready to present the action for the system of N nearly coincident  M$0$--branes (mM$0$ system) which was proposed in \cite{mM0=action}.
It can be considered as a result of  `gauging' of rigid  d=1 ${\cal N}=16$ supersymmetry (\ref{SYMsusy-X})--(\ref{SYMsusy-A}) of the $SU(N)$ SYM action with the Lagrangian (\ref{LSYM=1}) achieved by coupling it to a composite d=1 ${\cal N}=16$ supergravity (\ref{E++=Eu++}), (\ref{E+q=Ev+q}), (\ref{mM0:Om0=}) induced by embedding of the center of energy worldline of the mM$0$ system into the target 11D superspace.

The natural first step on this way is to make the Lagrangian (\ref{LSYM=1}) covariant by coupling it to a 1d gravity. This can be  reached by just replacing $d\tau$ in the right hand side of  (\ref{LSYM=1}) by  the 1-form $\hat{E}^{\#}$ of (\ref{E++=Eu++}). Then, to provide also the $SO(1,1)$ and $SO(9)$ gauge symmetries, which play  the role of Lorentz and R-symmetries in our induced 1d ${\cal N}=16$ supergravity, we should replace the Yang--Mills covariant derivatives in (\ref{SYMDX=}) by the $SO(1,1)\times SO(9)$ covariant derivatives defined in (\ref{DXi=}), (\ref{DPsi:=}), and to multiply the Lagrangian 1-form thus obtained by $(\rho^{\#})^3$. The next stage is suggested by the fact that setting N=1 in the action for the system of N nearly coincident M$0$ brane one should arrive a single M$0$-brane action.  As the SU(N) SYM Lagrangian, and all the matrix fields involved in it, vanish when N=1, this implies the necessity just to add the single M$0$ action to the integral of the above described Lagrangian form. Then the coupling to induced gravitino can be restored from the requirement of local supersymmetry invariance of the mM$0$ action.

\subsection{mM$0$ action}

In such a way we arrive at the mM$0$ action proposed in
\cite{mM0=action}. It reads
\begin{eqnarray}
\label{SmM0=} && S_{mM0} = \int_{W^1} \rho^{\#}\, \hat{E}^{=} + \qquad \nonumber  \\ &&
+ \int_{W^1} (\rho^{\#})^3\, \left(  tr\left(- {\bb P}^i D {\bb X}^i + 4i { \Psi}_q D
{\Psi}_q  \right) + \hat{E}^{\#} {\cal H} \right)+ \quad \nonumber \\ &&  +  \int_{W^1}
(\rho^{\#})^3\, \hat{E}^{+q}  tr\left(4i (\gamma^i {\Psi})_q  {\bb P}^i + {1\over 2}
(\gamma^{ij} {\Psi})_q  [{\bb X}^i, {\bb X}^j]  \right) , \; \nonumber \\ && {}
  \end{eqnarray}
where ${\cal H}$ is the relative motion Hamiltonian ({\it cf.} (\ref{HSYM=1}))
\begin{eqnarray}
\label{HmM0=} {\cal H} &:=& {\cal H}_{\#\#\#\#}({\bb X}, {\bb P}, \Psi ) = \qquad
\nonumber \\ &=& {1\over 2} tr\left( {\bb P}^i {\bb P}^i \right) + {\cal V} ({\bb X}) - 2\,  tr\left({\bb X}^i\, \Psi\gamma^i {\Psi}\right)  \qquad
  \end{eqnarray}
including  the scalar  potential ({\it cf.} (\ref{VSYM=}))
\begin{eqnarray}
\label{VmM0=} {\cal V} &:=& {\cal V}_{\#\#\#\#} ({\bb X} ) = - {1\over 64}
tr\left[ {\bb X}^i ,{\bb X}^j \right]^2 \qquad \\ &=& +{1\over 64}  tr\left[ {\bb X}^i
,{\bb X}^j \right] \cdot \left[ {\bb X}^i ,{\bb X}^j \right]^\dagger \; ,
  \end{eqnarray}
and  the Yukawa--type coupling  $tr\left({\bb X}^i\, \Psi\gamma^i {\Psi}\right)$.

The covariant derivatives $D$ are defined in (\ref{DXi=}), (\ref{DPsi:=}).
Their connection are build from the (spinor) moving frame variables, Eq.  (\ref{mM0:Om0=}), which are related to the center of energy motion of the mM$0$ system.  These  are also used to construct the composite graviton and gravitino 1-forms   $\hat{E}^{\#}$ and $\hat{E}^{+q}$, Eqs.  (\ref{E++=Eu++}), (\ref{E+q=Ev+q}). The 1-form $\hat{E}^{=}$ is the same as in the case of single M$0$-brane
\begin{eqnarray}
\label{E--=Eu--} \hat{E}^{=}=\hat{E}^{a}u_a^{=} \, . \qquad
  \end{eqnarray}
For the completeness of this section, let us recall that in these equations  $\hat{E}^{a}$ is the pull--back of the bosonic supervielbein to the center of energy worldline $W^1$, Eq. (\ref{hEa=}), (\ref{Ea=Pi}), $\hat{E}^\alpha
=d\hat{\theta}^\alpha(\tau)$, $u_a^{=}$ and $u_a^{\#}$ are light--like moving frame
vectors  (\ref{Uin}), (\ref{u--u--=0}), (\ref{u++u++=0}), and $v_\alpha^{+q}$ is an
element of spinor moving frame  (\ref{harmVin}).

Although the first term in (\ref{SmM0=}) coincides with the  single  M$0$--brane action
(\ref{SM0=}), now the Lagrange multiplier $\rho^{\#}$ and spinor moving frame variables are also present in
the second and third terms. This results in that their equations of motion differ from
(\ref{E==0=Ei}), and, as we discuss in the next section, generically, the center of energy motion of the mM0 system is not light-like.

\subsection{Local supersymmetry of the mM$0$ action}

The action (\ref{SmM0=}) is invariant under the transformation of the 16 parametric local worldline supersymmetry  \begin{eqnarray}
\label{susy-th} \delta_\epsilon \hat{\theta}^\alpha &=& \epsilon^{+q} (\tau)
v_q^{-\alpha} \; , \quad
 \\
\label{susy-x} \delta_\epsilon \hat{x}^a &=& - i \hat{\theta} \Gamma^a\delta_\epsilon
\hat{\theta} +   {1\over 2}   u^{a\#} i_\epsilon \hat{E}^{=}\; , \qquad
\\
\label{susy-rho}   \delta_\epsilon \rho^{\#} &=& 0\; , \qquad \\ \label{susy-v}
 \delta_\epsilon  v_q^{\pm\alpha}&=&0
  \;   \Rightarrow  \quad  \delta_\epsilon
  u_a^{=}= \delta_\epsilon u_a^{\#}= \delta_\epsilon u_a^{i}=0\;
 ,  \qquad
\\
\label{susy-X}  \delta_\epsilon {\bb X}^i   &=& 4i \epsilon^{+} \gamma^i  \Psi \; , \quad
\delta_\epsilon {\bb P}^i   = [(\epsilon^{+} \gamma^{ij}  \Psi),  {\bb X}^j]\; ,\qquad \\
\label{susy-Psi}  \delta_\epsilon \Psi_q &=&  {1\over 2} (\epsilon^{+} \gamma^i)_q  {\bb
P}^i-  {i\over 16} (\epsilon^{+} \gamma^{ij})_q  [{\bb X}^i, {\bb X}^j]\; ,\qquad \\
\label{susy-A}
 && \delta_\epsilon {\bb A} = -  \hat{E}^{\#} \epsilon^{+q}  \Psi_q + \hat{E}^{+}\gamma^i
 \epsilon^{+}\;    {\bb X}^i
 \; ,  \qquad
\end{eqnarray}
where
\begin{eqnarray}
\label{iE==}
& i_\epsilon \hat{E}^{=}= 6 (\rho^{\#})^2 tr \left(i {\bb P}^i\epsilon^{+}
 \gamma^{i}\Psi
 -  {1\over 8} \epsilon^{+} \gamma^{ij}\Psi  [{\bb X}^i, {\bb X}^j] \right) .   \qquad
 \end{eqnarray}

The local supersymmetry transformations of the  fields describing relative motion of mM$0$ constituents, (\ref{susy-X}), (\ref{susy-Psi})  coincide with the SYM supersymmetry (\ref{SYMsusy-X}), (\ref{SYMsusy-Psi}) modulo the fact that now the fermionic parameter is an arbitrary function of the center of energy proper time,  $\epsilon^{+q} =\epsilon^{+q}(\tau)$. The local supersymmetry transformation of the 1d $SU(N)$ gauge field (\ref{susy-A}) differs from the SYM transformation by additional term involving the composite gravitino.

The transformations of the center of energy variables Eqs. (\ref{susy-th})--(\ref{susy-v}) describe a deformation of the  irreducible  $\kappa$--symmetry (\ref{kap=irr}) of the free massless superparticle. Actually, the
deformation touches the  transformation rule (\ref{susy-x})  for the the bosonic
coordinate function,  $\delta_\epsilon \hat{x}^a$ only. The Lagrange multiplier $\rho^{\#}$ and the (spinor) moving  frame variables are invariant under the supersymmetry, like they are under the $\kappa$-symmetry of a single superparticle.

\section{${\bf m}$M0  equations of motion}
\label{SecEqmM0}
\setcounter{equation}{0}

In this section we derive  and study the equations of motion for the
multiple M$0$-brane system which follow from the action (\ref{SmM0=}).

\subsection{Equations of the relative motion}

Varying the action with respect to the momentum matrix field $\mathbb{P}^{i}$ gives us
the equation
\begin{eqnarray}\label{DXi=EPi}
&& D\mathbb{X}^i =\hat{E}^{\#}\mathbb{P}^i+4i\hat{E}^{+q}(\gamma^i\Psi)_q
 \qquad
\end{eqnarray}
which allows to identify $\mathbb{P}^{i}$, modulo fermionic contribution, with the
covariant time derivative of $\mathbb{X}^{i}$,
\begin{eqnarray}\label{Pi=DXi}
 \mathbb{P}^i=
D_\# \mathbb{X}^i - 4i\hat{E}_\#^{+}\gamma^i\Psi \; . \qquad
\end{eqnarray}
Here
\begin{eqnarray}\label{D++=d/E}
D_\# = {1\over \hat{E}_\tau^\#}D_\tau\; , \qquad  \hat{E}_\#^{+q} = {1\over
\hat{E}_\tau^\#}\hat{E}^{+q}_\tau\, , \qquad
\end{eqnarray}
are covariant derivative and the induced 1d gravitino field  corresponding to the induced einbein on the worldvolume,  $\hat{E}^\#=\hat{E}^au_a^\#=: d\tau \hat{E}^\#_\tau $, in the sense of that
\begin{eqnarray}\label{Pi=DXi}
D= \hat{E}^\# D_\# \; , \qquad \hat{E}^{+q}= \hat{E}^\# \hat{E}_\#^{+q}\; . \qquad
\end{eqnarray}
The variation with respect to the worldline gauge field ${\bb A}=d\tau {\bb A}_\tau $
gives
\begin{eqnarray}\label{Gauss}
[\mathbb{P}^i,\mathbb{X}^i]= 4i\{ \Psi_q \, , \, \Psi_q \}\, \qquad
\end{eqnarray}
and the variation with respect to ${\bb X}^i$ results in
\begin{eqnarray}\label{DPi=}
 D\mathbb{P}^i&=& -
\frac{1}{16}\hat{E}^{\#}[[\mathbb{X}^i,\mathbb{X}^j], \mathbb{X}^j] + \qquad \nonumber \\
&& +2\hat{E}^{\#}\, \Psi\gamma^{i}\Psi +\hat{E}^{+q}\gamma^{ij}_{qp}[\Psi_p,
\mathbb{X}^{j}]\, . \qquad
\end{eqnarray}
Using (\ref{DXi=EPi}) we can easily present this equation in the form
\begin{eqnarray}\label{DDXi=}
 D_\#D_\# \mathbb{X}^i&=& -
\frac{1}{16}[[\mathbb{X}^i,\mathbb{X}^j], \mathbb{X}^j] + 2 \Psi\gamma^{i}\Psi + \qquad
\nonumber \\ && + 4i
D_\#(\hat{E}_\#^{+}\gamma^{i}\Psi_p)+\hat{E}_\#^{+}\gamma^{ij}[\Psi, \mathbb{X}^{j}] \,
. \qquad
\end{eqnarray}
Finally, the variation with respect to the traceless matrix fermionic field  $\Psi_q$
produces
\begin{eqnarray}\label{DPsi=}
D\Psi &=& \frac{i}{4}\hat{E}^{\#}[\mathbb{X}^i,(\gamma^{i}\Psi)] + \qquad \nonumber \\
&& + \frac{1}{2} \hat{E}^{+}\gamma^{i}\, \mathbb{P}^{i}-\frac{i}{16}
\hat{E}^+\gamma^{ij} \, [\mathbb{X}^i,\mathbb{X}^j]\; .  \qquad
\end{eqnarray}

\subsection{A convenient gauge fixing}

To simplify the above equations,  let us use the fact that 1-dimensional connection can
always be gauged away and fix the gauge where the composed $SO(9)$ connection
(\ref{Omij:=}) and also the $SU(N)$ gauge field vanish
\begin{eqnarray}\label{Omij=0}
 \Omega^{ij}=d\tau \Omega_\tau^{ij}=0\; , \\
 \label{bbA=0}
  {\bb A}=d\tau {\bb A}_\tau =0 \; .
\end{eqnarray}
This breaks the local $SO(9)$ and $SU(N)$, but the symmetry under the rigid
$SO(9)\otimes SU(N)$ transformations remains.

As far as the $SO(1,1)$ gauge symmetry is concerned, we would not like to fix it but
rather use a part ${1\over 2}u^{a\#}\frac{\delta S_{mM0}}{{\delta} \hat{x}^{a}} =0$ of
the equations of motion for the center of energy coordinate functions $\hat{x}^{a}$
(discussed below in full),
\begin{eqnarray}\label{Drho=0}
&&D\rho^{\#}=0\qquad \; ,
\end{eqnarray}
to find the explicit form of the induced $SO(1,1)$ connection  (\ref{Om0:=}),
$\Omega^{(0)}:= {1\over 4} u^{a=}du_a^\#$. Indeed, as far as
$D\rho^{\#}=d\rho^{\#}-2\rho^{\#}\Omega^{(0)}$, Eq. (\ref{Drho=0}) implies
\begin{eqnarray}\label{Om0=}
 \Omega^{(0)}={d\rho^{\#}\over 2\rho^{\#}}\; .
\end{eqnarray}

In the gauge (\ref{Omij=0}), (\ref{bbA=0}) the set of bosonic gauge symmetries is
reduced to the Abelian $SO(1,1)$, $\tau$--reparametrization and $b$--symmetry (which we  describe below in sec. \ref{NoetherId}), and the
covariant derivatives simplify to
\begin{eqnarray}\label{D=dr}D\mathbb{X}^i&=& (\rho^{\#})^{-1}
d(\rho^{\#}\mathbb{X}^i)\, ,   \qquad \nonumber \\  D \mathbb{P}^i&=& (\rho^{\#})^{-2}
d((\rho^{\#})^2\mathbb{P}^i)\; , \qquad \nonumber \\ D\Psi_q &=&(\rho^{\#})^{-3/2}
d((\rho^{\#})^{3/2}\Psi_q)\; . \qquad \end{eqnarray}
As a result, Eqs.  (\ref{DDXi=}) and (\ref{DPsi=}) can be written in the following
(probably more transparent) form:
\begin{eqnarray}\label{dtPsi=}
\partial_\tau \tilde{\Psi} = \frac{i}{4} \,
e\,[\tilde{\mathbb{X}}{}^i,(\gamma^{i}\tilde{\Psi})]
+  \frac{1}{2\sqrt{\rho^{\#}}} \hat{E}_\tau ^{+}\gamma^{i}\, \tilde{\mathbb{P}}{}^{i} -
\qquad \nonumber \\ -\frac{i}{16\sqrt{\rho^{\#}}} \hat{E}_\tau^+\gamma^{ij} \,
[\tilde{\mathbb{X}}{}^i,\tilde{\mathbb{X}}{}^j]\;  , \qquad
\\ \label{ddXi=}
 \partial_\tau \left(\frac{1}{e}\partial_\tau \tilde{\mathbb{X}}{}^i\right) = -
\frac{e}{16}\, [[\tilde{\mathbb{X}}{}^i,\tilde{\mathbb{X}}{}^j],
\tilde{\mathbb{X}}{}^j] +2\, e \, \tilde{\Psi}\gamma^{i}\tilde{\Psi}  + \qquad
\nonumber \\
 +  4i \partial_\tau \left( { \hat{E}_\tau^{+}\gamma^{i}\tilde{\Psi}\over
 e\sqrt{\rho^{\#}}} \right) +
{1\over \sqrt{\rho^{\#}}} \hat{E}_\tau^{+}\gamma^{ij}[\tilde{\Psi},
\tilde{\mathbb{X}}{}^{j}]\, . \qquad
\end{eqnarray}
Writing  Eqs.  (\ref{dtPsi=}) and (\ref{ddXi=}) we used the redefined fields
\begin{eqnarray}\label{tX=rX}
\tilde{\mathbb{X}}{}^i&=&  \rho^{\#} {\mathbb{X}}{}^i\; , \qquad
\tilde{\Psi}_q=(\rho^{\#})^{3/2} {\Psi}_q\; , \qquad \\ \label{tP=rP}
\tilde{\mathbb{P}}{}^{i}&=&(\rho^{\#})^2{\mathbb{P}}{}^{i}= {1\over e}
\left(\partial_\tau \tilde{\mathbb{X}}{}^i - {4i\over \sqrt{\rho^{\#}}}
\hat{E}_\tau^{+}\gamma^i\tilde{\Psi}\right) \, , \quad
\end{eqnarray}
which are inert under  the $SO(1,1)$, and
\begin{eqnarray}\label{e=E/rho}
e(\tau)= \hat{E}^\#_\tau /\rho^{\#} \;  \quad
\end{eqnarray}
which has the properties of the einbein of the Brink--Schwarz superparticle action
(\ref{S'M0=}).

\subsection{By pass technical comment on derivation of the equations for the center of energy coordinate
functions}

This is the place to present some comments on the convenient way to derive equations of
motion for the center of energy variables (which was actually used as well when
working with single M$0$ in Sec \ref{M0eqs}). A reader not interested in technical
details may omit this subsection.

To find the manifestly covariant and supersymmetric invariant linear combinations of
the equations of motion for the bosonic and fermionic coordinate functions,
$\frac{\delta S_{mM0}}{\delta  \hat{x}^{a}}=0$ and $\frac{\delta S_{mM0}}{\delta
\hat{\theta}^{\alpha}}=0$, we introduce the covariant basis $i_\delta  \hat{E}^{A}$ in
the space of variation such that
\begin{eqnarray}\label{vhZS=}
 \delta_{\hat{Z}^M} S_{mM0} = \int_{W^1} \left(\delta  \hat{x}^{a} \frac{\delta
 S_{mM0}}{\delta  \hat{x}^{a}}+ \delta  \hat{\theta}^{\alpha} \frac{\delta
 S_{mM0}}{\delta  \hat{\theta}^{\alpha}}\right)=\qquad \nonumber \\
 = \int_{W^1} \left(i_\delta  \hat{E}^{a} \frac{\delta S_{mM0}}{i_\delta
\hat{E}^{a}}+ i_\delta  \hat{E}^{\alpha} \frac{\delta S_{mM0}}{i_\delta
\hat{E}^{\alpha}}\right)\; . \qquad
\end{eqnarray}
In  the generic case of curved superspace $i_\delta  \hat{E}^{A}= \delta \hat{Z}^{M}
{E}_M^{A}(\hat{Z})$; in our case of flat  target superspace this implies
\begin{eqnarray}\label{i-dEA=flat}
i_\delta  \hat{E}^{a}= \delta  \hat{x}^{a} - i \delta  \hat{\theta}\Gamma^a
\hat{\theta}\; , \qquad i_\delta  \hat{E}^{\alpha}= \delta  \hat{\theta}^{\alpha} \; .
\qquad
\end{eqnarray}
Furthermore, it is convenient to use the moving frame variables to split covariantly
the set of bosonic  equations $ \frac{\delta S_{mM0}}{i_\delta  \hat{E}^{a}}=0$  into
\begin{eqnarray}\label{vS/ivEb}
 \frac{\delta S_{mM0}}{i_\delta  \hat{E}^{=}}&=&{1\over 2}\, u^{a\#}\frac{\delta
 S_{mM0}}{i_\delta  \hat{E}^{a}}\, , \qquad
 \nonumber \\ \frac{\delta S_{mM0}}{i_\delta
 \hat{E}^{\#}}&=&{1\over 2}\, u^{a=}\frac{\delta S_{mM0}}{i_\delta  \hat{E}^{a}}\, , \qquad
 \nonumber \\
 \frac{\delta S_{mM0}}{i_\delta  \hat{E}^{i}}&=&- u^{a i}\frac{\delta S_{mM0}}{i_\delta
 \hat{E}^{a}}\; , \qquad
\end{eqnarray}
and the set of fermionic equations, $ \frac{\delta S_{mM0}}{i_\delta
\hat{E}^{\alpha}} = \frac{\delta S_{mM0}}{\delta\hat{\theta}^{\alpha}}$, into
\begin{eqnarray}\label{vS/ivEf}
\frac{\delta S_{mM0}}{i_\delta  \hat{E}^{-q}}=v_q^{+\alpha}\frac{\delta
S_{mM0}}{\delta  \hat{\theta}^{\alpha}} \; , \qquad \nonumber \\ \frac{\delta S_{mM0}}{i_\delta
\hat{E}^{+q}}=v_q^{-\alpha}\frac{\delta S_{mM0}}{\delta  \hat{\theta}^{\alpha}} \; .
\qquad
\end{eqnarray}
To resume,
\begin{eqnarray}\label{vhZS=cov}
 \delta_{\hat{Z}^M} S_{mM0} &=& \int\limits_{W^1} \, (\delta\hat{x}^{a}-i \delta
 \hat{\theta}\Gamma^a  \hat{\theta})  \left(u_a^=  \frac{\delta S_{mM0}}{i_\delta
 \hat{E}^{=}} + \right. \qquad \nonumber \\ && \left.
 +u_a^\#   \frac{\delta S_{mM0}}{i_\delta  \hat{E}^{\# }} + u_a^i  \frac{\delta
 S_{mM0}}{i_\delta  \hat{E}^{i}}
 \right)+ \nonumber \qquad\\
&&+ \int\limits_{W^1} \delta  \hat{\theta}^{\alpha} \left(v_\alpha^{-q} \frac{\delta S_{mM0}}{i_\delta  \hat{E}^{-q}}+ v_\alpha^{+q}  \frac{\delta S_{mM0}}{i_\delta  \hat{E}^{+q}}\right) . \nonumber \;\\ {} \qquad
\end{eqnarray}

\subsection{Equations for the center of energy coordinate functions}

As we  have already stated, the bosonic equation $\frac{\delta S_{mM0}}{i_{\delta}
\hat{E}^{=}}:= {1\over 2}u^{a\#}\frac{\delta S_{mM0}}{{\delta} \hat{x}^{a}} =0$ results
in Eq. (\ref{Drho=0}) which  is equivalent to (\ref{Om0=}). This observation is useful
to extract consequences of the next equation, $\frac{\delta S_{mM0}}{i_{\delta}
\hat{E}^{\#}}=0$, which  reads
\begin{eqnarray}\label{Dr3H=0}
&& D((\rho^{\#})^3 {\cal H})=0\; .
\end{eqnarray}
 Using (\ref{Drho=0}) one can write Eq. (\ref{Dr3H=0}) in the form of
 \begin{eqnarray}\label{dr4H=0}
&&d((\rho^{\#})^4 {\cal H})=0\; , \qquad
\end{eqnarray}
or, equivalently, $(\rho^{\#})^4 {\cal H}=const$. Due to the structure of ${\cal H}$, Eq. (\ref{HmM0=}),  this constant is nonnegative. Furthermore, as it has been shown in \cite{mM0=action} (see also sec. VIIIC),
it can be identified (up to numerical multiplier) with the mass parameter $M^2$ characterizing
the  center of energy motion,
\begin{eqnarray}\label{r4H=M2/4}
M^2= 4(\rho^{\#})^4 {\cal H}=const \geq 0 \; .
\end{eqnarray}

The remaining projection of the equation for the bosonic  center of energy
coordinate functions,  $\frac{\delta S_{mM0}}{i_{\delta} \hat{E}^{i}}:= - {1\over 2}u^{ai}\frac{\delta S_{mM0}}{{\delta} \hat{x}^{a}} =0$,  gives us the
relation between covariant ${SO(1,9)\over SO(1,1)\times SO(9)}$ Cartan forms
(\ref{Om++i=}),
\begin{eqnarray}\label{Om--=HOm++}
&&\Omega^{=i}=- (\rho^{\#})^2 {\cal H} \; \Omega^{\#i}= - {M^2\over 4(\rho^{\#})^2}\;
\Omega^{\#i}\;.  \qquad
\end{eqnarray}

The nontrivial part of the fermionic equation of the center of energy motion, $
\frac{\delta S_{mM0}}{i_{\delta} \hat{E}^{-q}}:= v_{q}^{-\alpha}\frac{\delta
S_{mM0}}{i_{\delta} \hat{E}^{\alpha}}=0$, reads
\begin{eqnarray}\label{E-q=Om}
&& \hat{E}^{-q}= -{1\over 2} \, \Omega^{\# i}\, \gamma^i_{qp}\nu_{\# p}^{\; -} \; ,
\end{eqnarray}
where
\begin{eqnarray}\label{nuiq=} \nu_{\# q}^{\; - } :=
(\rho^{\#})^{2}tr \left((\gamma^j \Psi)_q
\mathbb{P}^j-\frac{i}{8}(\gamma^{jk}\Psi)_q[\mathbb{X}^j,\mathbb{X}^k]\right)\, . \quad
\end{eqnarray}

\subsection{Noether identities for gauge symmetries. First look. }
\label{NoetherId}
Actually one can show that Eq. (\ref{Dr3H=0}) is satisfied identically when other
equations are taken into  account. (To be precise, Eqs. (\ref{DXi=EPi}), (\ref{Gauss}),
(\ref{DPi=}), (\ref{DPsi=}), (\ref{Drho=0}) have to be used). This is the Noether
identity for the 'tangent space' copy of the reparametrization symmetry (sometimes it
is called {\it 'b-symmetry'}) with the parameter function $i_\delta \hat{E}^\#$. Similarly, one can find the Noether identity reflecting the
existence of the ${\cal N}=16$ 1d gauge supersymmetry (\ref{susy-th})--(\ref{iE==}) with the basic parameter $\epsilon^{+q}=i_\delta \hat{E}^{+q}$ .
It states the dependence of the one half of the fermionic equations, namely
$\frac{\delta S_{mM0}}{i_{\delta} \hat{E}^{+q}}:= v_{q}^{-\alpha}\frac{\delta
S_{mM0}}{i_{\delta} \hat{E}^{\alpha}}=0$, which reads
\begin{eqnarray}\label{HE+q=}
D\nu_{\# q}^{\; - } =  {\rho}^2 \hat{E}^{+q}  {\cal H}\;
\end{eqnarray}
or $D\nu_{\# q}^{\; - } =  {\rho^{\#}}^2 \hat{E}^{+q}  {\cal H}_{\#\#\#\#}$ in a more complete notation.

\subsection{Equations which follow from the auxiliary field variations and simplification of the above
 equations}

Variation with respect to the Lagrange multiplier $\rho^{\#}$, $\frac{\delta
S_{mM0}}{\delta \rho^{\#}}=0$, express the projection $\hat{E}^{=}:= \hat{E}^au_a^{=}$
of the pull--back $\hat{E}^a$ of the bosonic supervielbein to the center of energy
worldline through the relative motion variables,
\begin{eqnarray}\label{E==mM0}
 \hat{E}^{=}&:=& \hat{E}^au_a^{=}= - 3 (\rho^\#)^2{\cal L}_{\#\#\# } =
\qquad \nonumber \\ & =&  3 (\rho^\#)^2 tr\left( \frac{1}{2}\mathbb{P}^iD\mathbb{X}^i +
\frac{1}{64} \hat{E}^{\#}[\mathbb{X}^i,\mathbb{X}^j]^2 - \right.  \qquad \nonumber \\
&&\left.  - \frac{1}{4}(E^+\gamma^{ij}\Psi)[\mathbb{X}^i,\mathbb{X}^j] \right)\; .
\qquad
\end{eqnarray}
The  $\hat{E}^{i}:= \hat{E}^au_a^{i}$  projection of this pull--back is expressed by
equations appearing as a result of variation with respect to the spinor moving frame
variables. According to Eqs. (\ref{vu++=iOm})--(\ref{vv+q}), those should appear as
coefficients for $i_\delta \Omega^{= i}$ and $i_\delta \Omega^{\# i}$ in the variation
of the action. Equation $\frac{\delta
S_{mM0}}{i_{\delta} \Omega^{= i}}=0$ reads
\begin{eqnarray}\label{Ei=mM0}
&&\hat{E}^{i}:= \hat{E}^au_a^{i}= - (\rho^{\#})^{-1}\, \Omega^{\# j} \, \left(J^{ij} +
\delta^{ij} J\right)\; ,  \qquad
\end{eqnarray}
where we have introduced the notation
\begin{eqnarray}\label{Jij=}
J^{ij}:= (\rho^{\#})^{3}\, tr\left( \mathbb{P}^{[i}\mathbb{X}^{j]}- i
\Psi\gamma^{ij}\Psi\right)\; ,\qquad \\ \label{J0=} J:= {(\rho^{\#})^{3}\,\over 2}
tr\left( \mathbb{P}^{i}\mathbb{X}^{i}\right)\; .  \qquad
\end{eqnarray}
The $(\rho^{\#})^{3}$ multipliers are introduced to make $J^{ij}$  and $J$ inert under
the $SO(1,1)$ transformations.

In this notation, equation $\frac{\delta S_{mM0}}{i_{\delta} \Omega^{\# i}}=0$ reads
\begin{eqnarray}\label{HEi=}
(\rho^\#)^3 {\cal H} \hat{E}^{i} =  -  \Omega^{= j} \left(J^{ij} - \delta^{ij} J\right)
- 2i(\rho^\#)  \hat{E}^{-q}(\gamma^{i}\nu_{\#}^{ -})_q\, . \nonumber \\ {}
\end{eqnarray}
Using (\ref{Om--=HOm++}), (\ref{Ei=mM0}), (\ref{r4H=M2/4}) and (\ref{E-q=Om}), one can
rewrite Eq. (\ref{HEi=}) as equation for $\Omega^{\# i}$,
\begin{eqnarray}\label{OmM2J=ff}
 \Omega^{\#  j} \, \left(M^2 J^{ij}
- 2i(\rho^\#)^3 \nu_{\#}^{-}\gamma^{ij}\nu_{\#}^{-} \right)=0 \, . \;\;
\end{eqnarray}
Actually, as we are going to show in the next sec. \ref{G}, {\it taking into account the remnant of the $K_9$ gauge symmetry of single M$0$-brane} (see (\ref{vK9}) and  (\ref{vK9v})) , which is present in the mM$0$ action, {\it one  can  present the above equation in the form of}
\begin{eqnarray}\label{Om++i=0}
 \Omega^{\#  i} \,  =0 \, , \qquad
\end{eqnarray}
 or, in terms of component, $\Omega_\tau^{\#  j}=0$.
 Due to (\ref{Om--=HOm++})  Eq. (\ref{Om++i=0}) implies
\begin{eqnarray}\label{mM0:Om--i=0}
 \Omega^{=  i} \,  =0 \,  \qquad
\end{eqnarray}
and (\ref{Ei=mM0}) acquires the same form as in the case of single M$0$-brane,
\begin{eqnarray}\label{hEi=0mM0}
&&\hat{E}^{i}:= \hat{E}^au_a^{i}= 0\; .  \qquad
\end{eqnarray}
Furthermore, the fermionic equation of motion (\ref{E-q=Om}) also becomes homogeneous, of the same form as the equation for  single M$0$-brane,
\begin{eqnarray}\label{E-q=0mM0}
&& \hat{E}^{-q}= 0 \; .
\end{eqnarray}
Eqs. (\ref{Om++i=0}) and (\ref{mM0:Om--i=0}) also  imply that all the moving frame and spinor moving frame variables are covariantly constant,
\begin{eqnarray}\label{mM0:Du=0}
Du_a^{\#}= 0\; ,  \qquad Du_a^{=}= 0\; ,  \qquad Du_a^{i}= 0\; ,  \qquad \\
\label{mM0:Dv=0}
Dv_q^{+\alpha}= 0\; ,  \qquad Dv_q^{-\alpha}= 0\; .  \qquad
\end{eqnarray}
Notice that in the case of single M$0$ brane such a form of equations for moving frame variables can be reached after gauge fixing the $K_9$ gauge symmetry with parameter $i_\delta \Omega^{\# i}$. In the mM$0$ case only a part (remnant) of $K_9$ symmetry is present so that a part of variations $i_\delta \Omega^{\# i}$ produce nontrivial equations which, together with the above mentioned remnant of $K_9$ symmetry, results in   Eqs. (\ref{mM0:Du=0}), (\ref{mM0:Dv=0}).

\subsection{Noether identity, remnant of the $K_9$ gauge symmetry and the final form of the $\Omega^{\# i}$ equation}

\label{G}

In this section we present the remnant of $K_9$ gauge symmetry leaving invariant the mM$0$ action  and show that, modulo this gauge symmetry, Eq. (\ref{OmM2J=ff}) is equivalent to (\ref{Om++i=0}).

Let us write  Eq. (\ref{OmM2J=ff}) as
\begin{eqnarray}\label{OmbbJ=0}
   \gimel^{ij} \Omega_\tau^{\#  j}= 0 \; , \qquad \end{eqnarray}
where \begin{eqnarray}\label{bbJ:=}
 \gimel^{ij}=M^2 J^{ij}
- 2i(\rho^\#)^3 \nu_{\#}^{-}\gamma^{ij}\nu_{\#}^{-} \, . \;\;
\end{eqnarray}
As this 9$\times$9 matrix is antisymmetric, it has rank 8 or lower, $rank (\gimel^{ij}) \leq 8$.
In other words, it has at least one `null--vector', this is to say a vector $V^i$ which obeys\footnote{This should not be confused with light--like vectors which can exist in the space with indefinite metric. In particular, our 11D moving frame vectors $u_a^=$ and $u^\#_a$ are light--like. To exclude any confusion, in this paper we never use the name 'null-vectors' for the  light--like vectors. }
\begin{eqnarray}\label{VbbJ=0}
\exists \; V^i, i=1,...,9 \; : \qquad  \gimel^{ij}V^j=0\, . \;\;
\end{eqnarray}
Actually, the matrix $\gimel^{ij}$ is constructed from the dynamical variables of our model
in  such a way (according to  Eqs.  (\ref{bbJ:=}) and (\ref{Jij=})) that  the number of its null vectors depends on the configuration of the fields describing the relative motion of the mM$0$ constituents.  However, as one `null vector' always exists, it is sufficient to consider a configuration with $rank (\gimel^{ij}) = 8$, and $\gimel^{ij}$ having just one `null vector',  at some neighborhood $\Delta \tau$  of a proper-time moment $\tau$; the generalization for a more complicated configurations/neighborhoods is straightforward.

Then, on one hand, the solution of  Eq. (\ref{OmbbJ=0}) in the neighborhood   $\Delta \tau$ is given by
$\Omega_\tau^{\# i}\propto V^i$, or, equivalently,
 \begin{eqnarray}\label{Om++=fv}
\Omega_\tau^{\# i}= f\, V^i\; , \qquad
\end{eqnarray}
where $f=f(\tau)$ is an arbitrary function of the center of energy proper time $\tau$. [For configurations/neighborhoods with several  `null vectors' $V^i_r$, $r=1,..., (9-rank \, \gimel)$ the solution will be $\Omega_\tau^{\# i}= f^r\, V_r^i$ with arbitrary functions $f^r=f^r(\tau)$].

On the other hand, the existence of null vector, Eq. (\ref{VbbJ=0}), implies that a part of Eqs. (\ref{OmbbJ=0}) is satisfied identically
\begin{eqnarray}\label{OmbbJV=0}
 \Omega_\tau^{\#  i} \, \gimel^{ij} V^j \equiv  0 \; , \qquad
\end{eqnarray}
when some other equations are taken into account. This is the {\it Noether identity} reflecting the existence of the gauge symmetry with the basic variation\footnote{See sec. \ref{bosonic}  for more details on these Noether identity and gauge symmetry in the purely bosonic case. Here let us just recall that Eq. (\ref{OmM2J=ff}) appears as an essential part of the coefficient for $i_\delta \Omega^{\# i}$ in the variation of the mM$0$ action.}
\begin{eqnarray}\label{iOm++i=aV}
i_\delta \Omega^{\# i}= \alpha \,  V^i\;  \qquad
\end{eqnarray}
with an arbitrary function $\alpha =\alpha (\tau )$.
This is clearly a remnant of the $K_9$ gauge symmetry (\ref{vK9}) of the action (\ref{SM0=}) for single M$0$-brane.

The generic variation of the Cartan 1--form $\Omega^{\# i}$ can be expressed as in Eq.  (\ref{vOm++=}),
which in our  1d case can also be written as
\begin{eqnarray}\label{vOm=DiOm}
\delta \Omega_\tau ^{\# i} = D_\tau i_\delta \Omega^{\# i} \;  . \qquad
\end{eqnarray}
Applying (\ref{vOm=DiOm}) to the variation of the solution (\ref{Om++=fv}) of Eq. (\ref{OmbbJ=0}) under (\ref{iOm++i=aV}), we find that
\begin{eqnarray}\label{vf=da}
\delta f(\tau)= \partial_\tau \alpha (\tau)\;  .   \qquad
\end{eqnarray}
Hence, one can use the local symmetry (\ref{iOm++i=aV}) to set $f=0$ and, thus, to gauge away (to trivialize) the solution (\ref{Om++=fv}) of Eq. (\ref{OmbbJ=0}).

This proves that  the gauge fixing version of Eq. (\ref{OmbbJ=0}) is given by Eq. (\ref{Om++i=0}), $\Omega^{\# i}=0$.

In sec. \ref{bosonic} we give more detailed discussion of the above local symmetry and its Noether identities reproducing independently the above conclusion for the purely bosonic case.

\section{Ground state solution of the relative motion  equations}
\label{grStSec}
\setcounter{equation}{0}

The natural first step in studying the above obtained mM$0$ equations is to address the
sector of
\begin{eqnarray}\label{Psiq=0}
\Psi_q=0 \, .  \;
\end{eqnarray}
As far as the fermionic equations of motion  have the same form (\ref{E-q=0mM0}) as for the single M$0$-brane, $\hat{E}^{-}_{q}=0$, the only possible fermionic contribution to the relative motion equations might come from
the induced gravitino $\hat{E}^{+q}=d\hat{\theta}^{\alpha}v_\alpha^{+q}$. However,
 with (\ref{Psiq=0}), the fermionic equation of the relative motion
(\ref{DPsi=}) results in
 \begin{eqnarray}\label{Psi=0=>E+}
 \hat{E}^{+}\gamma^{i}\, \mathbb{P}^{i}-\frac{i}{8} \hat{E}^+\gamma^{ij} \,
 [\mathbb{X}^i,\mathbb{X}^j]=0
\; .
\end{eqnarray}
As it will be clear after our discussion below, for $M^2>0$ this equation has only
trivial solution $\hat{E}^{+q}=0$, while for $M^2=0$ the 1d gravitino $\hat{E}^{+q}$
remains arbitrary.

\subsection{Ground state of the relative motion}

It is easy to see that a particular configuration of the bosonic fields for which Eq. (\ref{Psi=0=>E+}) is satisfied is
\begin{eqnarray}\label{grSt}
\mathbb{P}^{i}=0 \; , \qquad  [\mathbb{X}^i,\mathbb{X}^j]=0 \; .
\end{eqnarray}
Then the fermionic 1-form
$\hat{E}^{+q}$ remains arbitrary (and pure gauge) as it is in the case of single M$0$
brane.

Together with (\ref{Psiq=0}), Eqs. (\ref{grSt}) describe the ground state of the
relative motion. For it the relative motion Hamiltonian (\ref{HmM0=}) and the center of
energy effective mass vanish,
\begin{eqnarray}\label{M2=0mM0}
M^2=0\;
\end{eqnarray}
so that the center of energy motion is light--like. Moreover, when Eqs. (\ref{Psiq=0})
and (\ref{grSt}) hold, all the equations of the center of energy motion  coincide with
the equations for single M$0$--brane.

The ground state of the mM$0$ system is thus described by Eqs. (\ref{Psiq=0}),
(\ref{grSt}) and by a (pure bosonic) ground state solution of the single M$0$
equations. This preserves all 16 worldline supersymmetries, which corresponds (as we
have discussed in Sec. II) to the preservation of 16 of 32 spacetime supersymmetries.

\subsection{Solutions with $M^2=0$ have relative motion in the ground state sector}
\label{0M2=vacuum}

Curiously enough, being in the ground state of the relative motion is the only
possibility for the mM$0$ system to have the light--like center of energy motion
characterized by zero effective mass
\begin{eqnarray}
\label{M2=0} M^2=0 \; \Leftrightarrow \; {\cal H}= {1\over 2}tr ({\bb P}^i{\bb P}^i) -
{1\over 64} tr [{\bb X}^i, {\bb X}^j]^2 =0\; . \qquad
\end{eqnarray}
Indeed, the pure bosonic relative motion Hamiltonian ${\cal H}$ is given by the sum of
two terms both of which are traces of squares of hermitian operators ($[[{\bb X}^i,
{\bb X}^j]^\dagger=  [{\bb X}^j, {\bb X}^i]=- [{\bb X}^i, {\bb X}^j]$); hence, the sum
vanishes, ${\cal H}=0$, iff both equations in (\ref{grSt}) hold \footnote{We do not
discuss here the possible nilpotent contributions, like the possibility to solve the
equation  $a^2=0$ for a real bosonic $a(\tau)$ by
 $a=\beta_{\alpha_1...\alpha_{17}} \hat{\theta}^{\alpha_1}\ldots
 \hat{\theta}^{\alpha_{17}}$ with  17 center of energy fermions
 $\hat{\theta}^{\alpha}(\tau)$ contracted with  some fermionic
 $\beta_{\alpha_1...\alpha_{17}}= \beta_{[\alpha_1...\alpha_{17}]}$.}, ${\bb P}^i= D_\#
 {\bb X}^i=0$ and $[{\bb X}^i, {\bb X}^j]=0$ \footnote{This is true for finite size
 matrices. In the $N\mapsto \infty$ limit (mM$0$ condensate) one can  consider a
 'non--commutative plane' solution with $[{\bb X}^i, {\bb X}^j]=i\Theta^{ij}$  and
 c-number valued $\Theta^{ij}=-\Theta^{ji}$, see for instance, \cite{GZ+=JHEP11}. In
 the case of finite $N$ this solution cannot be used as far as the right hand side is
 assumed to be proportional to the unity matrix, $I_{N\times N}$ while the trace of the
 commutator vanishes. }.

Thus any nontrivial configuration of the relative motion, with either ${\bb P}^i\not=0$
or/and $[{\bb X}^i, {\bb X}^j]\not=0$,  creates a nonzero effective mass of the center
of energy motion, $M^2=0$.

\section{Supersymmetric solutions of  ${\bf m}$M0 equations}
\label{susySolSec}
\setcounter{equation}{0}

In this section we will show that  supersymmetric solutions of the mM$0$ equations have vanishing effective center of energy mass,  $M^2=0$, and that they can preserve only 1/2 of the target space supersymmetry. The first statement, which is tantamount to saying that mM$0$ BPS states are massless, can be found in \cite{mM0=action}, while the second, which is tantamount to saying that  mM$0$ BPS states are 1/2 BPS, is a new result of this paper.

\subsection{Supersymmetric solutions of the mM$0$ equations have $M^2=0$}

\label{BPS-M2=0}

From Eq. (\ref{susy-Psi}) one concludes that a solution of the mM$0$ equations with
vanishing relative motion fermionic fields, Eq. (\ref{Psiq=0}), can be supersymmetric
if
\begin{eqnarray}
\label{susy-Psi=0}
 (\epsilon^{+} \gamma^i)_q  {\bb P}^i-  {i\over 8} (\epsilon^{+} \gamma^{ij})_q  [{\bb
 X}^i, {\bb X}^j]=0\; . \qquad
\end{eqnarray}
All the 16 worldline supersymmetries (1/2 of the target space supersymmetries) can be
preserved iff this equation is satisfied for arbitrary $\epsilon^{+p}$. This implies
\begin{eqnarray}
\label{susy16-Psi=0} \gamma^i_{qp}  {\bb P}^i-  {i\over 8} \gamma^{ij}_{qp}  [{\bb
X}^i, {\bb X}^j]=0\;  \qquad
\end{eqnarray}
the only solution of which is given by the ground state of the relative motion, Eq.
(\ref{grSt}).

Thus all the bosonic solutions of mM$0$ equations preserving $16$ supersymmetries have
the trivial relative motion sector described by Eq. (\ref{grSt}) which is characterized
by the light--like center of energy motion, $M^2=0$.

This suggests that  $M^2=0$,  is the BPS condition,
{\it i.e.} the necessary condition for the 1/2 supersymmetry preservation. As we are going to show, this is indeed the case, and, moreover  \begin{eqnarray}
\label{M2=0BPS} M^2=0  \qquad
\end{eqnarray}
is the BPS equation for preservation of any part of the target space supersymmetry.

Indeed, on one hand, tracing Eq. (\ref{susy-Psi=0}) with $\gamma^j{\bb P}^j$ and using
the properties of $tr$ we find $$\epsilon^{+q}  tr({\bb P}^i{\bb P}^i)=  {i\over 8}
(\epsilon^{+} \gamma^{jk} \gamma^{i})_q  tr({\bb P}^i[{\bb X}^j, {\bb X}^k])\; . $$ On the
other hand, tracing (\ref{susy-Psi=0}) with ${i\over 8}\gamma^{jk}[{\bb X}^j, {\bb
X}^k]$ and using  the Jacobi identities $[{\bb X}^{[i}[{\bb X}^j, {\bb X}^{k]}]]\equiv
0$ we find $$ {i\over 8} (\epsilon^{+} \gamma^{i}\gamma^{jk})_q  tr({\bb P}^i[{\bb X}^j,
{\bb X}^k])= {1\over 32} (\epsilon^{+q} tr([{\bb X}^j, {\bb X}^k]^2)\;. $$ Taking the sum
of these two equations and using (\ref{Gauss}) (with fermionic fields set to zero) we find
$\epsilon^{+q}{\cal H}=0$  which, using (\ref{r4H=M2/4}), can be
written as $\epsilon^{+q} \, M^2 =0$,
\begin{eqnarray}
\label{BPS=0} \epsilon^{+q} \, M^2 =0\qquad \Leftarrow \qquad \epsilon^{+q}{\cal H}=0\;
.   \qquad
\end{eqnarray}
For $M^2\not=0 $ this implies $\epsilon^{+q}=0 $, so that the supersymmetry is broken.
Thus all the supersymmetric solutions of mM$0$ equation are characterized by $M^2=0$.

This fact is very important: it means that the existence of our action does not imply
the existence of a  new type  of supersymmetric solutions of the  11D SUGRA
equations\footnote{Although this statement can be done about the solutions preserving
1/2 of the 11D supersymmetry, as it will be clear in a moment, it is universal as far
as a supersymmetric solution of mM$0$ equations can preserve only 1/2 of the tangent
space supersymmetry.}. A BPS solution is in correspondence with the ground state of the
brane or of the multiple brane system; the ground state of mM$0$ system is
characterized by the vanishing effective mass and with the center of energy motion
characteristic for the single M$0$--brane. Thus a supersymmetric solution of 11D SUGRA equations  corresponding to  single M-wave also describe the mM$0$ (multiple M-wave) ground state.

\subsection{All BPS states of  mM$0$ system are 1/2 BPS}
\label{BPS=1/2}

As we have shown,  a solution of mM$0$ equations can preserve some part of the 16 worldline
supersymmetries (and some part ($\leq 1/2$) of the target space supersymmetry) if and
only if $M^2=0$. Now, in the light of the observation in sec. \ref{0M2=vacuum}, $M^2=0$
implies that the relative motion of the mM$0$ constituents is in its ground state, Eq.
(\ref{grSt}). This has two consequences. Firstly, as the ground state trivially solves the Killing spinor equation (\ref{susy-Psi=0}), it preserves all the supersymmetries allowed by the center of energy motion.
Secondly, when the relative motion sector is in its ground state, the center of energy sector of supersymmetric solution is described by the same equations as the motion of single M$0$--brane (massless 11D superparticle). Now, as we have shown in sec. \ref{susySOL}, the supersymmetric solutions of these M$0$ equations preserve just 1/2 of the target space supersymmetry.

{\it This proves that all the supersymmetric solutions of the equations of motion of the mM$0$  system preserve just one half of 32 target space supersymmetries. In other words, all the mM$0$ BPS states are 1/2 BPS.}

\section{On  solutions of $\mathbf m$M$0$ equations with $M^2> 0$ }
\label{bosonic}
\setcounter{equation}{0}

When $M^2\not=0$, Eq. (\ref{Psi=0=>E+}) has only trivial solutions. (The proof of this fact follows the stages of sec. \ref{BPS-M2=0}).  This means that  (\ref{Psiq=0}) results in
\begin{eqnarray}\label{E+q=0}
 \hat{E}^{+q}=0
\; ,
\end{eqnarray}
so that, when $M^2>0$, a configuration with vanishing relative motion fermion is purely bosonic.

\subsection{Purely bosonic equations in the case of $M^2>0$ }

The complete list of nontrivial pure bosonic equations for mM$0$ system with nonvanishing center of energy mass,
$M^2>0$, reads
\begin{eqnarray}\label{Drho0=0}
&& D\rho^{\#}=0\qquad \Leftrightarrow \quad   \Omega^{(0)}={d\rho^{\#}\over
2\rho^{\#}}\; ,
\\ \label{DDXi0=}
&& D_\#D_\# \mathbb{X}^i= - \frac{1}{16}[[\mathbb{X}^i,\mathbb{X}^j], \mathbb{X}^j] \; ,
\\
\label{Gauss0} && [D_\#\mathbb{X}^i,\mathbb{X}^i]= 0\, , \qquad
\qquad
\end{eqnarray}
\begin{eqnarray}\label{E0==mM0}
 \hat{E}^{=}&:=& d\hat{x}^{a}u_a^= =\qquad \nonumber \\
 &=& 3 \hat{E}^{\#}  \left(  (\rho^\#)^2  tr(D_\#\mathbb{X}^i)^2 -
 \frac{M^2}{4(\rho^\#)^2 } \right) ,  \\
\label{Ei0=mM0}
\hat{E}^{i}&:=& d\hat{x}^{a} u_a^{i}= 0 \, \; ,  \qquad
\end{eqnarray}
\begin{eqnarray}\label{OmM2J=0}
&& \Omega^{\#  i} \,=0 \, , \;\; \\
\label{Om--0=M2Om++}
&&\Omega^{=i}= 0\; ,  \qquad
\end{eqnarray}
where $\hat{E}^{\#}= d\hat{x}^{a}u_a^{\#}$ and the center of energy mass $M$ is defined by Eq. (\ref{r4H=M2/4}), $M^2= 4(\rho^{\#})^4
{\cal H}$, with the relative motion Hamiltonian
\begin{eqnarray}
\label{HmM00=} {\cal H}  &=& tr\left( {1\over 2}  (D_\# {\bb X}^i)^2   - {1\over 64}
\left[ {\bb X}^i ,{\bb X}^j \right]^2 \right)\; .
  \end{eqnarray}

Notice that (as we have discussed in the general case) the currents
\begin{eqnarray}\label{Jij0=}
J^{ij}&=& (\rho^{\#})^{3}\, trD_\#\mathbb{X}^{[i}\mathbb{X}^{j]}\; ,\qquad  \nonumber \\ J&=&
{(\rho^{\#})^{3}\,\over 2} tr D_\#\mathbb{X}^{i}\mathbb{X}^{i}\; \qquad
\end{eqnarray}
disappear from the final form of equations when one takes into account the presence of the remnants of the $K_9$ symmetry. As far as this statement is very important in the analysis of the mM$0$ equations, we are going to give more detail on this symmetry and gauge fixing now.

But before let us make an observation that  the current $J^{ij}$ is covariantly constant  on the mass shell (i.e. when the above equations of motion are taken into account),
\begin{eqnarray}\label{DJij0=0}
DJ^{ij}=0\; .
\end{eqnarray}
In contrast, in the generic purely bosonic configuration  the scalar current is not a constant, $DJ=dJ\not=0$.

\subsection{Remnant of $K_9$ symmetry in the  bosonic limit of the mM$0$ action and  $\Omega^{\#i}$ equations }

The variation of the bosonic limit of the mM$0$ action (\ref{SmM0=}) can be written in the form
\begin{eqnarray}\label{vSbos=}
\delta S^{^{bosonic}}_{mM0}&=& \int_{W^1} {\cal E}^{= i}_u i_\delta \Omega^{\# i}+ \int_{W^1} {\cal E}^{\# i}_u i_\delta \Omega^{= i}- \qquad \nonumber \\ && - \int_{W^1} {\cal E}_{\hat{x}}^{i} i_\delta \hat{E}^{i}+ \ldots  \; . \qquad \end{eqnarray}
where
 \begin{eqnarray}\label{cE=:=}
 {\cal E}^{= i}_u&=&M^2 \hat{E}^{i}/4 \rho^{\#} + \Omega^{= j}(J^{ij}-\delta^{ij}J)\; , \nonumber \\
  {\cal E}^{\# i}_u&=&\rho^{\#}  \hat{E}^{i}  + \Omega^{\# j}(J^{ij}+\delta^{ij}J)\; , \nonumber \\
{\cal E}_{\hat{x}}^{i}\; &=& \rho^{\#} \Omega^{= i}+ M^2 \Omega^{\# i}/4 \rho^{\#}\; , \qquad \end{eqnarray}
with $J^{ij}$ and $J$ defined in (\ref{Jij0=}) and  dots denote the terms involving the other basic variations ($\delta \rho^{\#}$, $i_\delta \hat{E}^=$ etc.). Furthermore, one can rearrange the terms in (\ref{vSbos=}) in the following way:
\begin{eqnarray}\label{vSbos==}
&& \delta S^{^{bosonic}}_{mM0}= \int_{W^1}  {\cal E}^{\# i}_u \left( i_\delta \Omega^{= i}-{ M^2\over 4(\rho^{\#})^2} i_\delta  \Omega^{\# i}\right)  - \; \nonumber
\\ && \qquad - \int_{W^1} {\cal E}_{\hat{x}}^{i} \left( i_\delta \hat{E}^{i}+ {1\over \rho^{\#}} \left(J^{ij}+\delta^{ij}J\right)\, i_\delta \Omega^{\# j} \right) +  \; \nonumber \\ && \qquad   + { M^2\over 2(\rho^{\#})^2}  \int_{W^1} d\tau\,  \Omega_\tau^{\# i} J^{ij} i_\delta \Omega^{\# j}  + \ldots
\, ,  \quad
\end{eqnarray}
In this form it is transparent that the equations of motion corresponding to the $i_\delta \Omega^{\# j}$ variation can be written in the form
\begin{eqnarray}\label{Om++iJij=0}
\Omega_\tau^{\# i} J^{ij} =0
\, ,  \quad
\end{eqnarray}
which is the bosonic limit of Eq.  (\ref{OmM2J=ff}). As we have already discussed in the general case, Eq. (\ref{Om++iJij=0}) always has a nontrivial solution as far as the antisymmetric $9\times 9$ matrix $J^{ij} = - J^{ji} $ always has at least one  null vector, a non-zero vector  $V^i$ such that
$V^iJ^{ij}=0$.

Each null--vector generates a nontrivial solution of (\ref{Om++iJij=0}), but also a gauge symmetry of the mM$0$ action. Indeed,  as one can easily see from (\ref{vSbos==}), the transformations with $\tau$-dependent parameter $i_\delta \Omega^{\# j}$ obeying
\begin{eqnarray}\label{JiOm++=0}
 J^{ij} i_\delta \Omega^{\# j} =0
\, ,  \quad
\end{eqnarray}
completed by
 \begin{eqnarray}\label{iOm--=iOm++}  i_\delta \Omega^{= i}&=& { M^2\over 4(\rho^{\#})^2} i_\delta  \Omega^{\# i} , \qquad \nonumber  \\
i_\delta \hat{E}^{i}\; &=&  - (J^{ij}+J\delta^{ij}) i_\delta \Omega^{\# j}/\rho^{\#}
\; , \qquad
\end{eqnarray}
leave the action invariant, $\delta S^{^{bosonic}}_{mM0}=0$, and, thus define the gauge symmetries of the mM$0$ action.
The transformations of  $\Omega_\tau^{\# i}$ under this gauge symmetry are $\delta \Omega_\tau^{\# i}=D_\tau i_\delta \Omega^{\# i}$ (\ref{vOm=DiOm}).  As far as in purely bosonic limit  $DJ^{ij}=0$ on the mass shell (see Eq. (\ref{DJij0=0})),
\begin{eqnarray}\label{JiOm++=0}
 J^{ij} D_\tau i_\delta \Omega^{\# j} =0\;    \quad
\end{eqnarray}
is also obeyed.  Furthermore, in 1d case all the connection can be gauged away so that the transformation rules of the nontrivial solution of Eq. (\ref{Om++iJij=0}) can be summarized as follows
\begin{eqnarray}\label{vOm++i=dtidOm++i}
\delta \Omega_\tau^{\# i}=\partial _\tau i_\delta \Omega^{\# i}\; , \qquad \begin{cases} \Omega_\tau^{\# i} J^{ij} =0 \, \cr  J^{ij} i_\delta \Omega^{\# j} =0 \; ,  \cr
\partial_\tau J^{ij}  =0 \; .
\end{cases}
\end{eqnarray}
This form makes transparent that any  nontrivial solution of Eq. (\ref{Om++iJij=0}) can be gauged away using local symmetry (\ref{JiOm++=0}), (\ref{iOm--=iOm++}). Thus, modulo the gauge symmetry,  Eq. (\ref{Om++iJij=0}) is equivalent to Eq. (\ref{OmM2J=0}), $\Omega^{\# i}=0$.

\subsection{Center of energy velocity and momentum for $M^2\not=0$}
\label{velocitySec}

Let us notice  one property of the center of energy motion of our M$0$ system which, on the first glance, might looks strange, and try to convince the reader that it is rather a natural manifestation of the  influence of relative motion on the center of energy dynamics.

Using Eqs.  (\ref{E0==mM0}), (\ref{Ei0=mM0}) we can easily calculate  center of energy velocity  of the bosonic limit of our mM$0$ system,
\begin{eqnarray}\label{dhx=mM00}
 && \dot{\hat{x}}^a:= \partial_\tau{\hat{x}}^a=  {1\over 2}\hat{E}_\tau^{=}u^{\# a}+ {1\over 2}\hat{E}_\tau^{\# }u^{= a}-\hat{E}_\tau^{i}u^{i a}= \nonumber \; \\
&&\;= {1\over 2} \hat{E}_\tau^{\# } \left(u^{= a}+ 3u^{\# a}\left(  (\rho^\#)^2  tr(D_\#\mathbb{X}^i)^2 -
 \frac{M^2}{4(\rho^\#)^2 } \right)\right)  .  \nonumber \\ {}
\end{eqnarray}
On the other hand, the canonical momentum conjugate to the center of energy coordinate function $\dot{\hat{x}}^a$ is\footnote{${\cal L}^{mM0}_\tau$ is the Lagrangian of the mM$0$ action (\ref{SmM0=}), $S_{mM0}=\int d\tau {\cal L}^{mM0}_\tau$.}
\begin{eqnarray}\label{p=mM0}
p_a= {\partial {\cal L}^{mM0}_\tau\over \partial \dot{x}^a}= \rho^{\#}\left( u_a^{=}+ u^{\#}_{a} \, \frac{M^2}{4(\rho^\# )^2} \right)\; .
\end{eqnarray}
This equation justifies our identification of the constant $M^2$ as a square of the effective mass of the mM$0$ system as it gives
\begin{eqnarray}\label{p2=M2}
p^ap_a=M^2  \; .
\end{eqnarray}
Thus, generically, the center of energy velocity and its momentum are oriented in different directions of 11D spacetime,
\begin{eqnarray}\label{velocity=p-cA}
\dot{\hat{x}}_a &\propto & \left( p_a - {\cal A}_a \right)\; , \qquad \\
\label{cA=}  && {\cal A}_a=  u^{\#}_{ a}\left( \frac{M^2}{\rho^\# } -3  (\rho^\#)^3  tr(D_\#\mathbb{X}^i)^2 \right)
\; . \qquad
\end{eqnarray}

Eq. (\ref{velocity=p-cA}) might look strange if one expects the center of energy motion to be similar to the motion of a free particle. However, this relation is characteristic for a charged particle moving in a background Maxwell field (see e.g. \cite{Landau}). In our case the counterpart  (\ref{cA=}) of the electromagnetic potential $ {\cal A}_a$ is constructed in terms of the relative motion variables. It vanishes when the relative motion is in its ground state.

Thus the seemingly unusual effect of that the mM$0$ center of energy  velocity and momentum are not parallel one to another is just one of the manifestations of the mutual influence of the center of energy and the relative motion in mM$0$ system. The relative motion variables, when they are not in ground state, generate a counterpart of the 11D background vector potential for the center of energy motion.

\subsection{An example of  non-supersymmetric solutions}

Let us fix the gauge (\ref{Omij=0}), (\ref{bbA=0}), ${\Omega^{ij}=0={\bb A}}$,  use the $SO(1,1)$ gauge symmetry to set  $\rho^{\#}=1$ and the reparametrization symmetry to fix $\hat{E}^{\#}_\tau=1$\footnote{Actually, to be precise,  there exists an obstruction to fix such a gauge by $\tau$ reparametrization  \cite{e=1}. The best what one can do is to fix $\partial_\tau \hat{E}^{\#}_\tau=0$, while the constant value remains indefinite. This is especially important for path integral quantization, where the integration over this constant value (mudulus)  should be included in the definition of the path integral measure.  As here we do not need in this level of precision, we allow ourselves to simplify the formulas by just setting this indefinite constant to unity. },
\begin{eqnarray}\label{gauge=}
\Omega_\tau ^{ij}=0={\bb A}_\tau \; , \qquad  \hat{E}^{\#}_\tau=1=\rho^{\#}\, . \quad
\end{eqnarray}
Then \begin{eqnarray}\label{D++=dt}
D_\#=\partial_\tau  \qquad
\end{eqnarray}
and  Eqs. (\ref{DDXi0=}) simplify to
\begin{eqnarray}\label{DDXi00g=}
\ddot{{\mathbb{X}}}{}^i&=& -
\frac{1}{16}\,
[[{\mathbb{X}}{}^i,{\mathbb{X}}{}^j]{\mathbb{X}}{}^j]\, , \quad
\\ \label{Gauss00}
&& [\dot{{\mathbb{X}}}{}^i,{\mathbb{X}}{}^i]= 0\, . \qquad
\end{eqnarray}
These very well known  equations describe the 1d reduction of the 10D $SU(N)$ Yang-Mills gauge theory.

A very simple solution of Eqs.  (\ref{DDXi00g=}) and (\ref{Gauss00}) is provided by
\begin{eqnarray}\label{X=At+B}
&&{\mathbb{X}}^i(\tau)=(A^i \tau +B^i){\mathbb{Y}},\qquad
\end{eqnarray}
where ${\mathbb{Y}}$ is a  constant traceless $N\times N$ matrix, $A^i$ and $B^i$ are constant  $SO(9)$ vectors, and $\tau$ is the proper time of the mM$0$ center of energy. The  center of energy effective mass is defined by the trace of ${\mathbb{Y}}^2$ and by the length of vector $\vec{A}=\{ A^i\}$,
\begin{eqnarray}
\label{M2=X2} M^2= 4{\cal H}  &=& 2 \vec{A}^2 tr {\mathbb Y}^2 \; , \qquad \vec{A}^2:=A^iA^i \; .
  \end{eqnarray}
Actually, by choosing the initial point of the proper time, $\tau \mapsto \tau - a$,
we can always make the constant SO(9) vectors $A^i$ and $B^i$ orthogonal,
\begin{eqnarray}\label{AB=0}
&& \vec{A}\vec{B}:= A^iB^i=0 .\qquad
\end{eqnarray}
Then the `currents' (\ref{Jij0=})  read
\begin{eqnarray}\label{Jij00=}
J^{ij}=
{A}^{[i}{B}^{j]} tr\mathbb{Y}^2= {{A}^{[i}{B}^{j]} \over 2\vec{A}^2} \; M^2 , \qquad
 J={\tau \over 4}\,  M^2\, .\quad
\end{eqnarray}

Now  the equations for the center of  energy coordinate functions (\ref{E==mM0}), (\ref{Ei0=mM0}) and the gauge fixing condition $\hat{E}_\tau^{\#}=1$ imply
\begin{eqnarray}\label{dx==mM0}
 \dot{\hat{x}}{}^{a}u_a^= &=&  {3M^2}/{4} , \qquad
\\ \label{dxi=mM0}
\dot{\hat{x}}{}^{a} u_a^{i}&=& 0  \, \; ,  \qquad
\\ \label{dx++=mM0}
 \dot{\hat{x}}{}^{a}u_a^\#  &=&1\, \; .  \qquad
\end{eqnarray}
With our gauge fixing, Eqs. (\ref{mM0:Du=0}), which follow from (\ref{OmM2J=0}), (\ref{Om--0=M2Om++}), implies that moving frame vectors are constant
\begin{eqnarray}\label{mM0:du=0}
\dot{u}_a^{\#}= 0\; ,  \qquad \dot{u}_a^{=}= 0\; ,  \qquad \dot{u}_a^{i}= 0\; . \qquad
\end{eqnarray}
Thus  (\ref{dx==mM0}), (\ref{dxi=mM0}), (\ref{dx++=mM0}) is a simple  system of linear differential equations
\begin{eqnarray}\label{dx===}
 && \dot{\hat{x}}{}^{=}   =  {3M^2}/{4} , \qquad
\\ \label{dxi==0}
&& \dot{\hat{x}}{}^{i}= 0  \, \; ,  \qquad
\\ \label{dx++==}
&& \dot{\hat{x}}{}^{\#}  =1\, \; ,  \qquad
\end{eqnarray}
for the variables
\begin{eqnarray} \label{x=anal}
\hat{x}^= =\hat{x}^a u_a^= \; , \qquad \hat{x}^\# =\hat{x}^a u_a^\# \; , \qquad \hat{x}^i =\hat{x}^a u_a^i \; .  \qquad
\end{eqnarray}
This system can be easily solved for the 'comoving frame' coordinate functions (\ref{x=anal}). The solution  in an arbitrary frame
\begin{eqnarray} \label{hx=sol1}
\hat{x}^\mu(\tau) =\hat{x}^\mu(0) + {\tau \over 2}\left(u^{= \mu}+ {3M^2\over 4} u^{\# \mu} \right)
\end{eqnarray}
describe a time-like motion of the center of energy characterized by a nonvanishing effective  mass (\ref{M2=X2}).
The velocity of this motion,
\begin{eqnarray} \label{dhx=sol1}
\dot{x}^\mu = {1\over 2}\left(u^{=\mu}+{ 3M^2\over 4} u^{\mu\#}\right)
\end{eqnarray}
is not parallel to the canonical momentum (see (\ref{p=mM0}))
\begin{eqnarray} \label{dhx=sol1}
p_\mu =  u^=_\mu+ {M^2\over 4} u_\mu^\# \; .
\end{eqnarray}
As it was discussed in general case in sec. \ref{velocitySec}, this is due to the influence of the relative motion of the mM$0$ constituents on the center of energy motion and can be considered as an effect of the induction by the relative motion dynamics of a counterpart of the Maxwell background field interacting with the center of energy coordinate functions. In the case under consideration this induced Maxwell field is  constant, ${\cal A}_\mu=- u_\mu^\#\, M^2/2 $.

\subsection{Another non-supersymmetric formal solution}

In the case of the system of 2 M$0$ branes, the $2\times 2$ matrix  field $\bb X^i$ can be decomposed on Pauli matrices, $\bb X^i= f^i_J(\tau) \sigma^J$,
 \begin{eqnarray} \label{sigmaI}
\sigma^I\sigma^J = \delta^{IJ} I_{2\times 2} + i\epsilon^{IJK}\sigma^K\; ,\qquad I,J,K=1,2,3. \qquad
\end{eqnarray}
The simplest ansatz which solves the Gauss constraint (\ref{Gauss00}) is $f^i_J(\tau) =\delta^i_Jf (\tau)$  so that \begin{eqnarray} \label{sigmaI}
{\bb X}^i (\tau) = f (\tau)\delta^i_J \sigma^J , \quad i=1,...,9; \;\;  I,J,K=1,2,3. \quad
\end{eqnarray}
Eq. (\ref{DDXi00g=}) then implies that this function should obey
\begin{eqnarray} \label{ddf=f3}
\ddot{f}+{1\over 2} f^3=0\, .\quad
\end{eqnarray}
The simplest solution of this equation is given by $f(\tau)={2i\over \tau}$ which is complex and thus breaks the condition that $\bb X^i$ is a hermitian matrix. Actually one can consider this solution,
\begin{eqnarray} \label{X=2i/tS}
{\bb X}^i (\tau) = {2i\over \tau}\, \delta^i_J \sigma^J , \quad J=1,2,3 .\quad
\, \quad
\end{eqnarray}
as an analog of instanton as far as the Wick rotation $\tau\mapsto i\tau$ restores the hermiticity properties.

Ignoring for a moment  the problem with hermiticity we can calculate the Hamiltonian and find that it is equal to zero. Thus (\ref{X=2i/tS}) is a solution with vanishing center of energy mass, $M^2=0$.

A configuration (\ref{sigmaI}) with nonzero effective center of energy mass can be obtained by observing that (\ref{ddf=f3}) has a more general solution given by  the so--called Jackobi elliptic function \cite{Abr}. These functions obey
\begin{eqnarray} \label{df=-f4+C}
\dot{f}^2= -f^4/4+ C
\, \quad
\end{eqnarray}
with an arbitrary constant $C$.  The above discussed particular solution (\ref{X=2i/tS}) of (\ref{ddf=f3}) solves (\ref{df=-f4+C}) with $C=0$ which suggests the relation of $C$ with $M^2$.  Indeed, a straightforward calculation shows that $C=M^2/12$  so that a solution of the mM0 equations of relative motion is given by 2x2 matrices (\ref{sigmaI}) with the function $f(\tau)$ obeying
\begin{eqnarray} \label{df=-f4+M2}
\dot{f}^2= {M^2-3f^4\over 12}\, .
\, \quad
\end{eqnarray}

The set of equations for  the center of energy motion includes  (\ref{dxi==0}), (\ref{dx++==}) and
\begin{eqnarray}\label{dx==c+f4}
 && \dot{\hat{x}}{}^{=}   =  {3M^2}/{4} - {9(f(\tau))^4}/{2} \; . \qquad
\end{eqnarray}
This equations can be solved numerically, but its detailed
study goes beyond the scope of this paper.

\section{Conclusions and discussion}

In this paper we obtain and study the equations of motion of multiple M0-brane
(multiple M-wave or shortly mM$0$)  system. In particular, we have shown that all the supersymmetric
solutions of mM$0$ equations preserve just one half of the 11D supersymmetry and are characterized by a trivial relative motion sector. This implies that all the  mM$0$ BPS states are 1/2 BPS and have the same properties as BPS states of single M$0$-brane. In the light of the possibility to describe the BPS states by the solution of the  supergravity field equations this implies that our results do not suggest existence of new exotic solutions of 11D supergravity: the mM$0$ BPS states are described by the same type supergravity solutions as the single M-wave
(see \cite{TomasBook} for discussion on this solution).

Our mM$0$ equations follow from the covariant supersymmetric and $\kappa$--symmetric mM$0$ action proposed in \cite{mM0=action} and we have also studied the gauge symmetries of these action. In particular we have found that this mM$0$ action is invariant under an interesting reminiscent of the so--called $K_9$ gauge symmetry characteristic for the spinor moving frame formulation of 11D massless superparticle (which is to say of single M$0$--brane). The accounting of this symmetry is necessary to find the final form of the bosonic equations of motion for the center of energy coordinate functions. This allows to check that the center of energy dynamics does not suffer indefiniteness, as might seem when looking on the original form of the center of energy equations which includes some number of arbitrary functions of proper time: just the above mentioned reminiscent of the $K_9$ symmetry allows to gauge away all these arbitrary functions.

Our equations for the system of N M$0$--branes are split on the equations for center of energy coordinate functions and moving frame variables, which are of the same type as the fields describing a single M$0$--brane, and the relative motion equations involving the bosonic and fermionic traceless $N\times N$ matrix fields
${\bb X}^i$ and ${\Psi}_q$ (as well as  auxiliary matrix fields: momentum  ${\bb P}^i$ and 1d $SU(N)$ gauge potential ${\bb A}$). The center of energy variables also enter the relative motion equations. There exists also the `backreaction'- the influence of the relative motion on the motion of the center of energy. This is characteristic for the purely bosonic Myers actions \cite{Myers:1999ps} and their generalizations \cite{YLozano+=0207}, but was not catched by the superembedding approach to mM$0$ system developed in \cite{mM0=PLB,mM0=PRL} because it was based on the standard superembedding approach equation for the center of energy variables. How to change this center of energy superembedding equation to account for `backreaction' of the relative motion on the center of energy dynamics is one of the interesting problems for future.

The most important effect of the `backreaction' of the relative motion (noticed already in \cite{mM0=action}) is that, in distinction to the case of a single M$0$--brane,  the generic center of energy motion of mM$0$ system is characterized by a nonvanishing effective mass $M$ constructed from the matrix field describing the relative motion.  Its square is expressed by $M^2= 4(\rho^{\#})^4 {\cal H}$ in terms of relative motion Hamiltonian ${\cal H}$ and the Lagrange multiplier $\rho^{\#}$ (which can be gauged to a constant). Both  $\rho^{\#}$ and ${\cal H}$  are covariantly constant on mass shell (i.e. when equations of motion are taken into account) and this guaranties that   $M^2$ is constant. The fact that this constant in nonnegative can be easily seen from the explicit expression for the relative motion hamiltonian  ${\cal H}$.

Another `backreaction' effect consists in that, when the relative motion is not in its ground state, the center of energy velocity and the  canonical momentum conjugate to the center of energy coordinate function are oriented in different directions of the 11D spacetime. This can be treated as an effect of  interaction of the center of energy coordinate degrees of freedom with the  counterpart of Maxwell background field induced by the relative motion.

All the `backreaction' effects disappear when $M^2=0$. In the purely bosonic case, it is easy to see (sec. \ref{0M2=vacuum}) that, when $M^2=0$, the relative motion is in its ground state described  by  constant commuting ${\bb X}^i$ matrices, Eqs. (\ref{grSt}). Moreover, we have found that $M^2=0$ is the BPS conditions for supersymmetric purely bosonic solutions of the mM$0$ equations (sec. \ref{BPS-M2=0}). This implies that all the supersymmetric bosonic solutions of the  mM$0$ equations   preserve just 1/2 of the target space supersymmetry (16 of 32), which implies that all the BPS states of mM$0$ system are 1/2 BPS. The proof uses, among the others,  the fact that all the BPS states of a single M$0$-brane are 1/2 BPS, which we have demonstrated in the introductory Sec II devoted to spinor moving frame formulation of a single M$0$ brane (11D massless superparticle) model.

Furthermore, we have shown that  all the supersymmetric solutions of mM$0$ equations have the relative motion sector in its ground state. For this the relative motion Hamiltonian vanishes ${\cal H}=0$, and, hence,  the effective mass of the center of energy motion of mM$0$ system is equal to zero, $M^2=0$. Then the center of energy momentum is light--like and parallel to the center of energy velocity. Moreover, all the equations of the center of energy motion acquire the same form as equations for single M$0$-brane , so that all the supersymmetric solutions of the mM$0$  equations are characterized by a solution of single M$0$-brane equations, describing the light--like movement of the center of energy of these supersymmetric mM$0$ configuration plus the nanoplet of constant commuting traceless $N\times N$ matrices ${\bb X}^i$ (where $N$ is the number of constituents of the mM$0$ system). These latter moduli of the mM$0$ system are the same as in 1d $SU(N)$ SYM theory.

One of the most important problems for future study is the search for generalization of our  mM$0$ action for the mM$0$ system in an arbitrary 11D supergravity background. Such a search does not promise to be simple (see
\cite{nonAbDp} for relevant studies of related bosonic models) so that  different approximations seems to be welcome. Probably a good starting point is to search for the generalization to  the case of curved superspace with constant fluxes, such as $AdS_{4(7)}\times S^{7(4)}$  and pp-wave superspaces\footnote{The purely bosonic Myers-type action for the mM0 system in a bosonic pp-wave background was proposed in  \cite{Lozano:2005kf}. The supersymmetric and Lorentz covariant equations for mM0 system in pp-wave superspace has been deduced in \cite{mM0-pp=PRD} from the superembedding approach description of
\cite{mM0=PRL}. This has been developed for an arbitrary 11D supergravity superspace, but  the dynamics of the center of energy in it has been  governed by the
the same superembedding equation as  describing single M$0$--brane. As a result, center of energy dynamics was considered to be not influenced by the relative motion, and plays a role of  background  for this. The present study of mM0 system in flat superspace shows that such an influence does exist, so that the center of energy superembedding equation has to be modified by the terms involving the fields describing relative motion of the mM0 constituents. Probably such a modified superembedding description of mM$0$ in pp-wave superspace, or the generalization of the flat superspace action of this paper to this case, may catch additional nonlinear terms in the mM$0$ equations, which are not present in \cite{mM0-pp=PRD}. The advantage of the action principle is that this distinguishes between three form potential $A_3$ of 11D supergravity and its dual $A_6$, so that its development for curved superspace might describe the interaction with $A_6$, similar to the one presented in   the purely bosonic  action of \cite{Lozano:2005kf}.}.

Another important problem is to understand whether it is possible to generalize our mM$0$ action for the case of multiple M$2$-brane (mM$2$) system. Both these problems are under investigation now.

\acknowledgments {\bf Acknowledgments}. {The authors are thankful to Dmitri Sorokin  for  useful discussions and to Juan Maria Aguirregabiria and Manuel Angel Valle for useful comments. This work was supported in part by the research grant FPA2012-35043-C02-01 from the MICINN (presently MEC) of Spain, by the Basque Government Research Group Grant ITT559-10 and by the UPV/EHU under the program UFI 11/55.}

\section*{APPENDIX A Equations of motion for a single M$0$ brane}
\renewcommand\theequation{A.\arabic{equation}}
\setcounter{equation}{0}

In this appendix we collect the equations of motion for the single M0--brane obtained from the spinor moving frame action (\ref{SM0=}), (\ref{SM0==}).  They read
\begin{eqnarray}
\hat{E}^{=}&:=& \hat{E}^{a}u_a^{=} = 0,\qquad \\
\hat{E}^{i}&:=& \hat{E}^{a}u_a^{i} =0, \qquad \\
 D\rho^{\#}&=&0 \quad \Leftrightarrow \quad \Omega^{(0)}={d\rho^{\#}\over 2\rho^{\#}}\; ,  \qquad \\
\Omega^{= i}&=&0 \quad \Leftrightarrow \quad Du_a^{=}=0 \quad \Leftrightarrow \quad \nonumber \\ && \qquad \Leftrightarrow \quad
Dv_q^{-\alpha}=0\, , \quad \\ \hat{E}^{-q}&:=& \hat{E}^{\alpha}v_\alpha^{-q}=0\, .
\end{eqnarray}
These equations are formulated in terms of pull--backs of bosonic and fermionic supervielbein forms of flat 11D superspace to the mM$0$ worldline $W^1$
\begin{eqnarray}\label{Ea=Pi-A}
 \hat{E}^a  = d\hat{x}^a - i d\hat{\theta} \Gamma^a\hat{\theta}\; , \qquad a=0,1,...,10\; , \\
E^\alpha=d\hat{\theta}^\alpha
 \;   \qquad \alpha= 1,...,32\; ,
  \end{eqnarray}
which are constructed from the coordinate functions $\hat{x}^a(\tau)$, $\hat{\theta}^\alpha (\tau)$ of the proper time $\tau$, and of the moving frame and spinor moving frame variables $u_b^{=}$, $u_b^{i}$, $v_\alpha^{-q}$. The properties of these latter as well as of the Cartan forms $\Omega^{=i}$, $\Omega^{(0)}$ and covariant derivatives $D$ are collected in the next Appendix B.

In (\ref{Ea=Pi-A}) and in the main text we have used the real symmetric $32\times 32$ 11D $\Gamma$--matrices $\Gamma^a_{\alpha\beta}= (\gamma^a C)_{\alpha\beta}$ which, together with  $\tilde{\Gamma}_a^{\alpha\beta}= (C\gamma_a)^{\alpha\beta}$, obey  $\Gamma^{(a}\tilde{\Gamma}^{b)}=\eta^{ab}{\bb I}_{32\times 32}$.

\section*{APPENDIX B Moving frame and spinor moving frame variables}
\renewcommand\theequation{B.\arabic{equation}}
\setcounter{equation}{0}

Moving frame and spinor moving frame variables are defined as blocks of, respectively, $SO(1,10)$ and $Spin(1,10)$ valued matrices,
\begin{eqnarray}\label{Uin-A}
& U_b^{(a)}= \left({u_b^{=}+ u_b^{\#}\over 2}, u_b^{i}, { u_b^{\#}-u_b^{=}\over 2}
\right)\; \in \; SO(1,10)\;  \quad
\end{eqnarray}
($i=1,...,9$) and    \begin{eqnarray}\label{harmVin-A} V_{(\beta)}^{\;\;\; \alpha}=
\left(\begin{matrix}  v^{+\alpha}_q
 \cr  v^{-\alpha}_q \end{matrix} \right) \in Spin(1,10)\;
 \; . \qquad
\end{eqnarray}
We also use
\begin{eqnarray}\label{Vharm=M0-A}
 V^{( {\beta})}_{ {\alpha}}= \left(
v_{ {\alpha}q}{}^+\, ,v_{ {\alpha}q}{}^- \right)\; \in \; Spin(1,10) \; ,  \qquad
\end{eqnarray}
with \begin{eqnarray}
\label{V-1=CV-A}  v_{\alpha}{}^{-}_q =  i C_{\alpha\beta}v_{q}^{- \beta }\, ,
\qquad v_{\alpha}{}^{+}_q = - i C_{\alpha\beta}v_{q}^{+ \beta }\,
 \end{eqnarray}
 obeying
 \begin{eqnarray}\label{Vharm=M0--A}
 V_{( {\beta})}{}^{ {\gamma}}
V_{ {\gamma}}^{ ({\alpha})}=\delta_{( {\beta})}{}^{ ({\alpha})}=\left(\begin{matrix}
\delta_{qp} & 0           \cr
          0 & \delta_{qp} \end{matrix}\right) \qquad  \\ \nonumber \Leftrightarrow  \quad \begin{cases} v_{q}^{- {\alpha}}v_{ {\alpha}p}{}^+=\delta_{qp}= v_{q}^{+
{\alpha}}v_{ {\alpha}p}{}^-\, , \cr  v_{q}^{- {\alpha}}v_{ {\alpha}p}{}^-= 0\; =
v_{q}^{+ {\alpha}}v_{ {\alpha}p}{}^+\, . \end{cases}\;
\end{eqnarray}

The algebraic properties of moving frame and spinor moving frame variables are summarized as
\begin{eqnarray}\label{u--u--=0-A}
u_{ {a}}^{=} u^{ {a}\; =}=0\; , \quad    u_{ {a}}^{=} u^{ {a}\,i}=0\; , \qquad u_{
{a}}^{\; = } u^{ {a} \#}= 2\; , \qquad
 \\  \label{u++u++=0-A} u_{ {a}}^{\# } u^{ {a} \#
}=0 \; , \qquad
 u_{{a}}^{\;\#} u^{ {a} i}=0\; , \qquad  \\  \label{uiuj=-A} u_{ {a}}^{ i}
 u^{{a}j}=-\delta^{ij}.  \qquad
\\
\label{M0:v+v+=u++-A}
 v_{q}^- {\Gamma}_{ {a}} v_{p}^- = \; u_{ {a}}^{=} \delta_{qp}\; , \qquad v_{q}^+ {\Gamma}_{ {a}} v_{p}^+ = \; u_{ {a}}^{\# } \delta_{qp}\; , \qquad \nonumber \\
 v_{q}^- {\Gamma}_{ {a}} v_{p}^+ = - u_{ {a}}^{i} \gamma^i_{qp}\; , \qquad
\\ \label{M0:u++G=v+v+-A}
  2 v_{q}^{- {\alpha}}v_{q}^{-}{}^{ {\beta}}= \tilde{\Gamma}^{ {a} {\alpha} {\beta}} u_{
 {a}}^{=}\; , \quad 2 v_{q}^{+ {\alpha}}v_{q}^{+}{}^{ {\beta}}= \tilde{\Gamma}^{ {a} {\alpha} {\beta}} u_{
 {a}}^{\# }\; , \qquad \nonumber \\
 2 v_{q}^{-( {\alpha}}v_{q}^{+}{}^{ {\beta})} =-  \tilde{\Gamma}^{ {a} {\alpha} {\beta}}
 u_{ {a}}^{i}\; . \qquad
\end{eqnarray}

In (\ref{M0:v+v+=u++-A}) and (\ref{M0:u++G=v+v+-A}) we have used  real symmetric $16\times 16$
9d Dirac matrices  $\gamma^i_{qp}=\gamma^i_{pq}$ which obey
Clifford algebra \begin{eqnarray}\label{gigj+=-A} \gamma^i\gamma^j + \gamma^j \gamma^i=
2\delta^{ij} I_{16\times 16}\; , \qquad
\end{eqnarray}
and
\begin{eqnarray}\label{gi=id1-A}
&& \gamma^{i}_{q(p_1}\gamma^{i}_{p_2p_3) }= \delta_{q(p_1}\delta_{p_2p_3) }\; , \qquad
\\ \label{gi=id2-A} && \gamma^{ij}_{q(q^\prime }\gamma^{i}_{p^\prime)p }+
\gamma^{ij}_{p(q^\prime }\gamma^{i}_{p^\prime)q } = \gamma^{j}_{q^\prime
p^\prime}\delta_{qp}-\delta_{q^\prime p^\prime}\gamma^{j}_{qp} \; . \qquad
\end{eqnarray}

Derivatives of the moving frame and spinor moving frame variables are expressed in terms of
covariant ${SO(1,10)\over SO(1,1)\times SO(9)}$ Cartan forms
\begin{eqnarray}
\label{Om++i=-A} \Omega^{=i}= u^{=a}du_a^{i}\; , \qquad \Omega^{\# i}=  u^{\#
a}du_a^{i}\; , \qquad
  \end{eqnarray}
and induced  $SO(1,1)\times SO(9)$ connection
\begin{eqnarray}
\label{Om0:=-A} \Omega^{(0)}= {1\over 4} u^{=a}du_a^{\#}\; , \qquad \\ \label{Omij:=-A}
\Omega^{ij}=  u^{ia}du_a^{j}\; . \qquad
  \end{eqnarray}
It is convenient to use these latter to define covariant derivative. Then
\begin{eqnarray}\label{M0:Du--=Om-A}
Du_{ {b}}{}^{=} &:= & du_{ {b}}{}^{=} +2 \Omega^{(0)} u_{ {b}}{}^{=}= u_{ {b}}{}^i
\Omega^{= i}\; , \qquad \\ \label{M0:Du++=Om-A} Du_{ {b}}{}^{\#}&:=& du_{ {b}}{}^{\#} -2
\Omega^{(0)} u_{ {b}}{}^{\#}=  u_{ {b}}{}^i \Omega^{\# i}\; , \qquad \\
\label{M0:Dui=Om-A}  Du_{ {b}}{}^i &:=& du_{ {b}}{}^{i} - \Omega^{ij} u_{ {b}}{}^{j} =
{1\over 2} \, u_{ {b}}{}^{\# } \Omega^{=i}+ {1\over 2} \, u_{ {b}}{}^{=} \Omega^{\#
i}\; . \qquad \nonumber \\ {}
\end{eqnarray}
\begin{eqnarray}
\label{Dv-q-A}  Dv_q^{-\alpha}&:=& dv_q^{-\alpha} +  \Omega^{(0)} v_q^{-\alpha} - {1\over
4}\Omega^{ij} \gamma^{ij}_{qp} v_p^{-\alpha} = \nonumber \\ &=& - {1\over 2}
\Omega^{=i} v_p^{+\alpha} \gamma_{pq}^{i}\; , \qquad \\ \label{Dv+q-A}  Dv_q^{+\alpha}
&:=& dv_q^{+\alpha} -  \Omega^{(0)} v_q^{+\alpha} - {1\over 4}\Omega^{ij}
\gamma^{ij}_{qp} v_p^{+\alpha} = \nonumber \\ &=& - {1\over 2} \Omega^{\# i}
v_p^{-\alpha} \gamma_{pq}^{i}\; . \qquad
\end{eqnarray}

The Cartan forms obey
\begin{eqnarray}\label{M0:DOm--=-A} && D\Omega^{= i}=  0\; , \qquad D\Omega^{\# i}  = 0\;  ,  \qquad
\\ \label{M0:Gauss-A}
 && F^{(0)}:= d\Omega^{(0)} =    {1\over 4 } \Omega^{=\, i} \wedge
 \Omega^{\# \, i}\; , \qquad  \\
\label{M0:Ricci-A} && {G}^{ij}:= d\Omega^{ij}+ \Omega^{ik} \wedge \Omega^{kj} = - \Omega^{=\,[i} \wedge \Omega^{\# \, j]}\; .   \qquad
\end{eqnarray}
Notice that, {\it e.g.}
 \begin{eqnarray}
\label{DDu++=} && DDu_{ {a}}^{\# } = - 2 F^{(0)}u_{ {a}}^{\#
}\; , \qquad
 DDu_{ {a}}{}^{i} = u_{ {a}}^{j} {G}^{ji}  \; . \qquad
\end{eqnarray}

The essential variations of moving frame and spinor moving frame variables can be written as
\begin{eqnarray}\label{vu--=iOm-A}
\delta u_{ {b}}{}^{=} = u_{ {b}}{}^i i_\delta\Omega^{= i}\; , \qquad \label{vu++=iOm-A}
\delta u_{ {b}}{}^{\#}=  u_{ {b}}{}^i i_\delta\Omega^{\# i}\; , \qquad \\
\label{vui=iOm-A}  \delta u_{ {b}}{}^i  = {1\over 2} \, u_{ {b}}{}^{\# }  i_\delta
\Omega^{=i}+ {1\over 2} \, u_{ {b}}{}^{=}  i_\delta\Omega^{\# i}\; . \qquad
\\
\label{vv-q-A}  \delta v_q^{-\alpha}= - {1\over 2} i_\delta \Omega^{=i} v_p^{+\alpha}
\gamma_{pq}^{i}\; , \qquad \\ \label{vv+q-A}  \delta v_q^{+\alpha} = - {1\over 2}
i_\delta\Omega^{\# i} v_p^{-\alpha} \gamma_{pq}^{i}\; , \qquad
\end{eqnarray}
where  $i_\delta
\Omega^{=i}$ and $i_\delta \Omega^{\# i}$ are independent variations.

The essential variations of the Cartan forms read
\begin{eqnarray}
\label{vOm++=-A} \delta \Omega^{\# i}&=& D i_\delta \Omega^{\# i}\; , \qquad  \delta \Omega^{=i}= D i_\delta \Omega^{=i}\; , \qquad
 \\
\label{vOmij=-A} \delta \Omega^{ij}\; &=& \Omega^{=[i} i_\delta \Omega^{\# j]} -  \Omega^{\# [i}i_\delta \Omega^{=j]}\; , \qquad
\\
\label{vOm0=-A}
\delta \Omega^{(0)}&=& \frac {1}{4} \Omega^{=i}i_\delta \Omega^{\# i} -
 \frac {1}{4}  \Omega^{\# i}i_\delta \Omega^{=i}  \; . \qquad
\end{eqnarray}

\section*{APPENDIX C mM0 equations of motion}
\renewcommand\theequation{C.\arabic{equation}}
\setcounter{equation}{0}

The  mM$0$ system, which is to say an interacting system of N nearly coincident M$0$-branes, is described in terms of center of energy variables, which similar to the variables of a single M$0$-brane described in Appendix A, and the traceless $N\times N$ matrices ${\bb X}^i$ ($i=1,...,9$), $\Psi_q$ ($q=1,...,16$). Our action includes also the auxiliary  $N\times N$ matrix fields: momentum ${\bb P}^i$ and the 1d SU(N) gauge field ${\bb A}_\tau$
(${\bb A}=d\tau {\bb A}_\tau$).

The complete list of equations of motion for the mM0 system splits naturally on the equations for the relative motion variables,
\begin{widetext}
\begin{eqnarray}\label{relEqm}
&& D\mathbb{X}^i =\hat{E}^{\#}\mathbb{P}^i+4i\hat{E}^{+q}(\gamma^i\Psi)_q,\nonumber\\
&&[\mathbb{P}^i,\mathbb{X}^i]= 4i\{ \Psi_q \, , \, \Psi_q \}, \nonumber\\
&&D\mathbb{P}^i= - \frac{1}{16}\hat{E}^{\#}[[\mathbb{X}^i,\mathbb{X}^j]\mathbb{X}^j]
 +2\hat{E}^{\#}\, \Psi\gamma^{i}\Psi
+\hat{E}^{+q}\gamma^{ij}_{qp}[\Psi_p, \mathbb{X}^{j}], \nonumber\\ && D\Psi =
\frac{i}{4} \hat{E}^{\#}[\mathbb{X}^i,(\gamma^{i}\Psi)] +
\frac{1}{2} \hat{E}^{+}\gamma^{i}\, \mathbb{P}^{i}-\frac{i}{16} \hat{E}^+\gamma^{ij} \,
[\mathbb{X}^i,\mathbb{X}^j] \,  \qquad
\end{eqnarray}
and the center of energy equations which can be considered as a deformation of the system of
equations for single M$0$ brane. After fixing the gauge under a reminiscent of the $K_9$ symmetry,
these equation read
\begin{eqnarray}\label{E==mM0-A}
 \hat{E}^{=}&:=& \hat{E}^au_a^{=}=
  3 (\rho^\#)^2 tr\left( \frac{1}{2}\mathbb{P}^iD\mathbb{X}^i +
\frac{1}{64} \hat{E}^{\#}[\mathbb{X}^i,\mathbb{X}^j]^2 -  \frac{1}{4}(E^+\gamma^{ij}\Psi)[\mathbb{X}^i,\mathbb{X}^j] \right)\; ,
\qquad \end{eqnarray}
 \end{widetext}
 \begin{eqnarray}
\label{Ei=0mM0-A}
\hat{E}^{i}&:=& \hat{E}^au_a^{i}= 0 \; ,   \\ \hat{E}^{-q}&:=& \hat{E}^{\alpha}v_\alpha^{-q}=0\, ,\\
\left. \begin{matrix} \Omega^{= i}=0 \cr
  \Omega^{\# i}=0
\end{matrix} \right\}
&& \Leftrightarrow \quad \left\{  \begin{matrix} Du_a^{=}=0,   \quad Du_a^{\#}=0, \cr   \quad Du_a^{i}=0\, , \cr
Dv_q^{-\alpha}=0\, , \quad D v_q^{+\alpha}=0\, , \end{matrix}\right.   \qquad  \\
 D\rho^{\#}&=&0 \quad \Leftrightarrow \quad \Omega^{(0)}={d\rho^{\#}\over 2\rho^{\#}}\; .  \qquad
\end{eqnarray}

As a consequence of the above equation the effective mass $M$ of the mM$0$ center of energy motion, \begin{eqnarray}\label{M2=-A}
M^2= 4 (\rho^{\#})^4 {\cal H}\; ,  \qquad
\end{eqnarray}
 is a constant
\begin{eqnarray}\label{dM2=0-A}
dM^2=0\; .  \qquad
\end{eqnarray}
Eq. (\ref{M2=-A}) expresses $M^2$ in terms of Lagrange multiplier  $\rho^{\#}$ and the relative motion Hamiltonian (\ref{HSYM=1}) \begin{eqnarray}
\label{HSYM=1-A}
{\cal H}=   {1\over 2} tr\left( {\bb P}^i {\bb P}^i \right) - {1\over 64}
tr\left[ {\bb X}^i ,{\bb X}^j \right]^2 - 2\,  tr\left({\bb X}^i\, \Psi\gamma^i {\Psi}\right) .  \quad
  \end{eqnarray}

The definition and properties of the  covariant derivatives of the spinor moving frame
variables and of the Cartan forms, described in the main text, are collected in   Appendix B

If we fix the gauge where the composed SO(9) connection and also the SU(N) gauge field
vanish, \begin{eqnarray}
&&\Omega^{ij}=d\tau \Omega_\tau^{ij}=0\; , \qquad {\bb A}=d\tau {\bb A}_\tau =0,
 \end{eqnarray}
the equations of relative motion and Eq. (\ref{E==mM0-A}) simplify to
\begin{widetext}
\begin{eqnarray}
 &&\partial_\tau
\tilde{\Psi} = \frac{i}{4} \, e\,[\tilde{\mathbb{X}}{}^i,(\gamma^{i}\tilde{\Psi})] +
\frac{1}{2\sqrt{\rho^{\#}}} \hat{E}_\tau ^{+}\gamma^{i}\, \tilde{\mathbb{P}}{}^{i} -
\frac{i}{16\sqrt{\rho^{\#}}} \hat{E}_\tau^+\gamma^{ij} \,
[\tilde{\mathbb{X}}{}^i,\tilde{\mathbb{X}}{}^j]\;  ,\nonumber\\ &&
\partial_\tau
\left(\frac{1}{e}\partial_\tau \tilde{\mathbb{X}}{}^i\right) = - \frac{e}{16}\,
[[\tilde{\mathbb{X}}{}^i,\tilde{\mathbb{X}}{}^j]\tilde{\mathbb{X}}{}^j] +2\, e \,
\tilde{\Psi}\gamma^{i}\tilde{\Psi}  +  4i \partial_\tau \left(
{ \hat{E}_\tau^{+}\gamma^{i}\tilde{\Psi}\over e\sqrt{\rho^{\#}}} \right) + {1\over
\sqrt{\rho^{\#}}} \hat{E}_\tau^{+}\gamma^{ij}[\tilde{\Psi}, \tilde{\mathbb{X}}{}^{j}]\,
, \nonumber\\ &&\partial_{\tau} \tilde{\mathbb{X}}{}^i= e\tilde{\mathbb{P}}{}^i +
{4i\over \sqrt{\rho^{\#}}}\left( \hat{E}_\tau^{+}\gamma^{i}\tilde{\Psi}\right)\, ,
\qquad [\tilde{\mathbb{P}}{}^i, \tilde{\mathbb{X}}{}^i]=4i\{\tilde{\Psi}_q,
\tilde{\Psi}_q\}\, , \nonumber\\
&&\rho^{\#} \hat{E}_\tau^{=}=3 tr \left({1\over 2}\tilde{\mathbb{P}}{}^i\partial_\tau
\tilde{\mathbb{X}}{}^i+ {1\over 64}e
[\tilde{\mathbb{X}}{}^i,\tilde{\mathbb{X}}{}^j]^2-{1\over
4\sqrt{\rho^{\#}}}\left(\hat{E}_\tau^{+}\gamma^{ij}\tilde{\Psi}\right)[\tilde{\mathbb{X}}{}^i,\tilde{\mathbb{X}}{}^j]\right)\,
 .
\end{eqnarray}
These equations are written in terms of  redefined fields,
\begin{eqnarray}\label{tX=rX}
\tilde{\mathbb{X}}{}^i&=&  \rho^{\#} {\mathbb{X}}{}^i\; , \qquad
\tilde{\Psi}_q=(\rho^{\#})^{3/2} {\Psi}_q\; , \qquad
\tilde{\mathbb{P}}{}^{i}=(\rho^{\#})^2{\mathbb{P}}{}^{i}= {1\over e}
\left(\partial_\tau \tilde{\mathbb{X}}{}^i - {4i\over \sqrt{\rho^{\#}}}
\hat{E}_\tau^{+}\gamma^i\tilde{\Psi}\right) \, ,
\end{eqnarray}
\end{widetext}  and
\begin{eqnarray}
e(\tau)= \hat{E}^\#_\tau /\rho^{\#} \;  .
\end{eqnarray}

\end{document}